\documentclass[twocolumn,american]{revtex4-2}
\usepackage[T1]{fontenc}
\usepackage[latin9]{inputenc}
\usepackage{xcolor}
\usepackage{babel}
\usepackage{amsmath}
\usepackage{amssymb}
\usepackage{graphicx}
\usepackage{wasysym}
\PassOptionsToPackage{normalem}{ulem}
\usepackage{ulem}
\usepackage[pdfusetitle,
 bookmarks=true,bookmarksnumbered=false,bookmarksopen=false,
 breaklinks=false,pdfborder={0 0 1},backref=false,colorlinks=false]
 {hyperref}

\makeatletter

\newcommand*\LyXZeroWidthSpace{\hspace{0pt}}
\providecommand{\tabularnewline}{\\}
\providecolor{lyxadded}{rgb}{0,0,1}
\providecolor{lyxdeleted}{rgb}{1,0,0}
\DeclareRobustCommand{\mklyxadded}[1]{\bgroup\color{lyxadded}{}#1\egroup}
\DeclareRobustCommand{\mklyxdeleted}[1]{\bgroup\color{lyxdeleted}\mklyxsout{#1}\egroup}
\DeclareRobustCommand{\mklyxsout}[1]{\ifx\\#1\else\sout{#1}\fi}
\makeatother

\begin{document}
\title{Nonlinear phase synchronization and the role of spacing in shell models}
\author{L. Manfredini, {\"O}. D. G\"urcan}
\affiliation{Laboratoire de Physique des Plasmas, CNRS, Ecole Polytechnique, Sorbonne
Universit\'e, Universit\'e Paris-Saclay, Observatoire de Paris,
F-91120 Palaiseau, France}
\begin{abstract}
A shell model can be considered as a self-similar chain of interacting
triads, where each triad can be interpreted as a nonlinear oscillator
that can be mapped to a spinning top. Investigating the relation between
phase dynamics and intermittency in such a chain of nonlinear oscillators,
it is found that synchronization is linked to increased energy transfer.
In particular, our results indicate that the observed systematic increase
of intermittency, as the shell spacing is decreased, is associated
with strong phase alignment among consecutive triadic phases, facilitating
the energy cascade. It is shown that while the overall level of synchronization
can be quantified using a Kuramoto order parameter for the global
phase coherence in the inertial range, a local, weighted Kuramoto
parameter can be used for the detection of burst-like events propagating
across shells in the inertial range. This novel analysis reveals how
locally phase-locked states are associated with the passage of extreme
events of energy flux. Applying this method to helical shell models
( i.e. for a class of helical interactions that couple the two helicities
in a non separable topology) reveals that a reduction in phase coherence
correlates with suppression of intermittency. When inverse cascade
scenarios are considered using two different shell models including
a non local helical shell model, and a local standard shell model
with a modified conservation law, it was shown that a particular phase
organization is needed in order to sustain the inverse energy cascade.
It was also observed that the PDFs of the triadic phases were peaked
in accordance with the basic considerations of the form of the flux,
which suggests that a triadic phase of $\pi/2$ and $-\pi/2$ maximizes
the forward and the inverse energy cascades respectively.
\end{abstract}
\maketitle

\section{Introduction}

One of the greatest challenges in modern physics, especially concerning
statistical dynamics, is characterizing the behavior of complex systems
far from thermodynamic equilibrium. These systems are not only of
fundamental intellectual interest, due to the rich complexity of the
dynamics that they exhibit, but also for their relevance in a wide
range of engineering and environmental applications.

Many natural systems exhibit non-Gaussian fluctuations across a broad
range of spatial and temporal scales, implying the presence of extreme
fluctuations or rare events, whose dynamical properties and statistical
signatures are essential to describe the overall behavior of the system.
Examples include seismic activity, financial market crashes and extreme
weather events \citep{bak:2002,gopikrishnan:1999,lovejoy:2013}. An
archetypical example of such complex, open system is the problem of
fully developed turbulence: an out of equilibrium state of fluid or
plasma flow, displaying a mixture of hierarchical self organization
and chaotic behavior across a large range of scales, maintained in
a statistically stationary state by continuous energy injection (e.g.
at large scales) and dissipation (e.g. at small scales).

Between forcing and dissipation, turbulent systems can develop an
inertial range, spanning many decades, exhibiting self-similar properties
heuristically captured by the Richardson cascade picture, and the
power law spectrum of Kolmogorov \citep{frisch}.

Since direct numerical simulations (DNS) of turbulent flows over a
wide range of scales were historically too costly, various reduced
models were developed in the past \citep{bohr_Jensen_Paladin_Vulpiani_1998}.
These models are capable of reproducing various statistical features
of the full system, such as the power law distribution of energy as
a function of scale, and the scaling of higher-order structure functions.
Among those, shell models \citep{biferale:03,ditlevsen:2010} stand
out in their simplicity and in their ability to mimic certain statistical
aspects of different types of turbulent systems including rotating
turbulence, passive scalar advection, thermal convection etc. \citep{hattori:04,jensen1992,kumar:2015}.
In the context of plasma physics, shell models have been used to describe
magnetohydrodynamic (MHD) turbulence \citep{plunian:2013,verdini:12},
and drift waves in fusion plasmas \citep{berionni:17,ghantous:15}.

Probably the most intriguing -- perhaps unexpected -- feature of
shell models, is their ability to produce the anomalous scaling of
the structure functions similar to fully developed turbulence \citep{kadanoff1995,ohkitani:89,lvov:98}.
This suggests that shell models can be a useful tool to capture the
statistics of extreme events \citep{lvov:2001,de-wit:2024:}, which
are manifested as burst like structures, related to finite time singularities
or instantons \citep{mailybaev:12,mailybaev:2013}, playing an important
role in shaping the underlying intermittency characteristics.

Shell models reduce the complexity of the nonlinear energy transfer
governed by triadic interactions drastically, while preserving the
key properties of the underlying system such as its scale invariance,
the quadratic nature of the nonlinearity and the conservation laws
- like Energy and Helicity in the case of three dimensional turbulence.
This extreme reduction in the number of degrees of freedom, is a consequence
of dividing the Fourier space into a set of geometrically spaced shells
and retaining the nonlinear interactions that involve only consecutive
shells, thus replacing the turbulent system with a one dimensional
chain of connected, self-similar triads.

In order to reintroduce some spatial structure in cascade models,
various more sophisticated constructions have been proposed, including
nested polyhedra \citep{gurcan:17}, logarithmic discretization (LDM)
\citep{gurcan:16a}, spiral chains \citep{gurcan:19} and logarithmic
lattices \citep{campolina:2018,campolina:2021,costa:2023}. Most of
these models are self-similar by construction, and maintain a constant
number of triads across different scales. However, a puzzling feature
of these spatially-extended ``shell'' models is the apparent suppression
of intermittency observed in their statistics. Similar loss of intermittency
has also been observed in numerical simulations of fractally-decimated
3D turbulence \citep{lanotte:15} even keeping an overwhelming majority
of the modes and dropping only a small percentage.

The effect of a fractal decimation of Fourier modes has been explored
also in simpler systems, such as the Burgers equation, where a disruption
of the average triadic phase coherence is correlated with a reduction
of the intermittency, normally associated with shock formation \citep{buzzicotti:2016,buzzicotti-phase:2016}.
This suggests a picture, in which phase alignment between triads enhances
energy transfer resulting in strong dissipation events \citep{murray:2018,protas:2024}.
Extending this analysis to the Navier Stokes equations requires understanding
the self-organization of triadic phases in fully developed turbulence,
and the definition of a clear order parameter that identifies the
interactions involved in the collective behavior responsible for intermittency,
which remains a challenge. Part of the difficulty involves dealing
with the huge number of degrees of freedom represented by the convolution
sum in such a system. One possible way forward may be to reformulate
turbulence as a network of Fourier modes \citep{gurcan:20,gurcan:21,gurcan:23},
which allows us to consider reduced models, such as shell models,
as reductions of a complete grid into a network of groups of nodes,
in this case represented by shells. Although the individual complex
phases of the dynamical variables of a shell model have no physical
significance, the fact that the triadic phase coherence along the
chain regulates the cascade process, suggests that triadic phase coherence
among the underlying Fourier modes, lumped together to form the shells,
is related to the resulting shell-to-shell energy transfer.

Thus, to clarify the relationship between intermittency and phase
coherence in fully developed turbulence, we investigate this connection
in shell models that preserve the same invariants as the Navier-Stokes
equations and, in some cases, exhibit a level of intermittency comparable
to that of 3D turbulence. Building on our earlier observation that
reducing inter-shell spacing in standard shell models increases intermittency
considerably \citep{manfredini:2025}, we demonstrate a systematic
link between triadic phase synchronization (sync) and enhanced intermittency
in shell models, consistently with the observations in studies of
the Burger's equation cited above.

Notice that, a shell model can be interpreted as a one dimensional
chain of connected triads, each representing a ``nonlinear oscillator''.
Consecutive triads only differ by a scaling factor related to the
inter-shell spacing. This means that by decreasing the inter-shell
spacing, (while keeping the conserved quantities unchanged) both the
number and the similarity of the oscillators would be increased making
it potentially easier to form a phase coherent state. To quantify
global phase coherence, and its relation to different levels of anomalous
scaling of the structure functions, we employ a global Kuramoto order
parameter, which allows us to show that higher intermittency correlates
with enhanced phase coherence.

In addition, we observe that phase coherence correlates with the existence
of coherent structures in the energy cascade, appearing as localized
pulses in shell models. In order to clarify the role of these localized
events, we introduce a novel local order parameter --a weighted Kuramoto
parameter-- which enable us to observe burst-like events propagating
across shells in the inertial range and analyze the alignment of triadic
phases in forming an ``open channel'' for energy transfer during
these events. We show that extreme energy bursts are consistently
preceded by rapid growth in local phase coherence and that more intermittent
models display stronger and longer-lasting phase-locked states, thereby
facilitating larger energy transfer.

Beyond the usual linear chain, we also analyzed helical shell models,
which introduce the complexity of chiral interactions, fundamental
in describing transitions in turbulent cascades \citep{alexakis:18}.
Helical shell models consist in two sequences of helical modes that,
depending on the interaction class, may be topologically intertwined
by the triads that connect them. In a particular interaction class,
a reduction of the inter-shell spacing leads to a disruption of the
intermittency, offering us a framework to test our diagnostic tools
for quantifying phase coherence in this opposite scenario.

Finally, we consider inverse cascades in shell models, which -apart
from particular examples- are usually suppressed by equipartition
solutions. We consider these special cases that actually work, in
order to study the role of phase coherence and to compare it with
the forward cascade scenario. The results reveal that, in the forward
and inverse cascade regimes, the PDFs of the triadic phases exhibit
peaks at values of $\pi/2$ and $-\pi/2$ respectively, which maximize
the flux term at fixed amplitude. Notably, in the case of intermittent
cascades the peak at $\pi/2$ is much more marked, indicating that
coherent events are clearly favored by the system; in contrast, for
inverse cascades, the distribution is symmetric. This results are
in parallel and complementary with recent results focusing on the
relation between phase dynamics and the energy cascade direction \citep{benavides:2025}.

The rest of the paper is organized as follows.

In section \ref{sec:Shell-model-as} we introduce the shell model
framework, focusing on the dynamical equations for phases and amplitudes
\ref{subsec:Basic-Phase-Dynamics}, discussing role of phase alignments
in the energy flux \ref{subsec:Choice-of-phase-1} and presenting
the phase statistics for the standard forward intermittent cascade.
Next in \ref{subsec:Sync-and-Kuramoto-1}, we introduce the Kuramoto
order parameters to quantify phase coherence in shell models, highlighting
how local sync is related to extreme flux events.

Section \ref{sec:Intermittency-1} addresses intermittency, in particular
investigating how the enhancement of the intermittency corrections,
by decreasing the shell spacing $g$, is associated with different
statistics of the sync order parameters \ref{subsec:Shell-spacing-Dependence}.
We further characterize the dynamics of burst events - responsible
for intermittency - as sync events propagating through shells in cascade
process.

In Sec. \ref{sec:Helical-shell-models}, we extend the analysis to
helical shell models, exploring how sync properties vary with shell
spacing, focusing on a helical shell model exhibiting suppression
of intermittency. Section \ref{sec:Inverse-cascades} is devoted to
the role of phase organization in driving inverse cascades. In particular
we present an inverse energy cascade for both an elongated helical
shell model and a local shell model.

Finally, Section \ref{sec:Conclusion} summarizes conclusions, while
Appendix \ref{sec:Numerical-details} provides numerical details.

\section{\protect\label{sec:Shell-model-as}Shell model as a dynamical system:
equations for phases and amplitudes}

Motivated by observations of statistical isotropy and scale invariance
of the turbulent cascade of fully developed turbulence without obvious
sources of anisotropy, the shell model reduction consists in modeling
the velocity field as a one-dimensional set of complex variables representing
different scales, corresponding to a set of logarithmically spaced
wave numbers $k_{n}=k_{0}g^{n}$, where it is customary to take the
spacing factor $g=2$.

The evolution equation for the shell variable $u_{n}$, for the GOY
(Gledzer-Ohkitani-Yamada) model, which is the standard example for
shell models, is usually written as

\begin{align}
\frac{du_{n}}{dt} & =i\left[ak_{n+1}u_{n+2}^{*}u_{n+1}^{*}+bk_{n}u_{n-1}^{*}u_{n+1}^{*}+ck_{n-1}u_{n-1}^{*}u_{n-2}^{*}\right]\nonumber \\
+ & d_{n}u_{n}+f_{n}\;\text{,}\label{eq:goy}
\end{align}
where the terms in the square brackets represent the reduction of
the convolution sum to a local triadic interaction between three consecutive
shells, $d_{n}=\nu k_{n}^{2p}+\mu k_{n}^{-2q}$ is the general form
of (hyper/hypo)-viscosity, and $f_{n}$ denotes external forcing,
localized at a shell $n_{f}$, or a small set of consecutive shells
around it, corresponding to the typical energy injection scale $1/k_{n_{f}}$.

Here the coefficients $a,b,c$ are chosen in such a way that they
preserve typical inviscid invariants of the underlying physical system.
In the case of 3D Navier-Stokes turbulence, it is the energy and the
helicity that are conserved, which are defined for the shell models
respectively as

\begin{equation}
E=\sum_{n}|u_{n}|^{2},\quad H=\sum_{n}(-1)^{n}k_{n}|u_{n}|^{2}\;\text{.}\label{eq:inv_conserv_NS-1-1}
\end{equation}

Note the peculiar aspect of the GOY model and its variants that positive
and negative helicities correspond to even and odd shells respectively.

Enforcing the conservation of these two quantities one obtains, $a=g+1$,
$b=g^{-1}-g$ and $c=-1-g$, as a function of the inter-shell spacing
$g$. It is also common to factor out $ak_{n+1}$ in the nonlinear
term of Eq. (\ref{eq:goy}) so that the other two coefficients corresponding
to $b$ and $c$ can be written as $\left(g^{-2}-g^{-1}\right)$ and
$-g^{-3}$ respectively. The common choice of $g=2$ would then give
$-1/4$ and $-1/8$ for the second and third coefficients respectively.
Another common choice in the literature is $g=\varphi$ ( the golden
ratio), which gains significance, from the derivation of shell models
based on a decimation of Fourier space, which requires the existence
of triadic interations between the characteristic wave numbers of
the shells (specifically solution to $\boldsymbol{k}_{n-1}+\boldsymbol{k}_{n}+\boldsymbol{k}_{n+1}=0$,
which imposes the geometrical constrain $g\leq\varphi$ ). While there
are some examples of studies of shell models varying their nonlinear
coupling coefficients \citep{biferale:95,ditlevsen:1996}, most of
these studies are carried out in such a way that varying the coefficients
change the conservation law, and hence the effective dimensionality
of the underlying physical model, instead of keeping the conservation
laws fixed and varying the logarithmic spacing for a fixed model.
Exploring different shell spacings is also important for making the
connection to a perspective based on Fourier space decimation, such
as log-lattices \citep{campolina:2021}, LDM \citep{gurcan:16a},
or helical shell models with different types of elongations \citep{rathmann}.

Note that a standard shell model with $N$ nodes has $N-2$ interacting
triads. Since the equation for a single unconnected triad is equivalent
to that of a spinning top, giving rise to nonlinear oscillations that
can be represented in terms of Jacobi elliptic functions, each triad
can be considered as a nonlinear oscillator. Furthermore since the
shell model consists of self-similar triads that are scaled by the
same factor, the consecutive nonlinear oscillators that are coupled
are also self-similar, maintaining the same scaling factor. The energy
flux through such a chain of triads then requires some sort of synchronization
of the phases of those consecutive triads. Decreasing the shell spacing
is thus key to understanding the synchronization behavior of shell
models. Since it simultaneously allows us to increase $N$, considering
the large $N$ limit, important for synchronization, and reduce the
ratio of the coefficients of two consecutive triads allowing them
to have closer natural nonlinear frequencies. Note that the usual
GOY model with $g=2$, covers roughly seven decades in $k$-space
with $N=24$, and it is not feasible to increase that number substantially
without decreasing $g$, since the power law solution would start
going under machine precision levels.

Thus, varying the $g$ parameter, while increasing $N$ in order to
cover the same wave-number range, allows us to modify the affinity
of the triadic network of the shell model to synchronization (i.e.
\emph{synchronizability}). Since the intermittency in shell models
is a consequence of its time evolution, one would expect this to have
significant impact on intermittency as well. Hence, we present below,
a detailed study introducing specific methods for observing various
aspects of local and global synchronization, and the implications
of these observations for the turbulent cascade as well as intermittency,
mainly by varying the $g$ parameter to control its \textit{sychronizability}.

\subsection{\protect\label{subsec:Basic-Phase-Dynamics}Phase Dynamics in Shell
Models}

In order to clarify the relevance of phase organization, we substitute
$u_{n}=U_{n}e^{i\theta_{n}}$ in Eq. (\ref{eq:goy}), and separate
the system into an equation for the amplitudes

\begin{equation}
\begin{aligned}\frac{dU_{n}}{dt}= & ak_{n+1}U_{n+2}U_{n+1}\sin\varphi_{n+1}+bk_{n}U_{n+1}U_{n-1}\sin\varphi_{n}\\
 & +ck_{n-1}U_{n-1}U_{n-2}\sin\varphi_{n-1}\\
 & +d_{n}U_{n}+Re\left(f_{n}e^{-i\theta_{n}}\right)
\end{aligned}
\label{eq:amp}
\end{equation}
and a corresponding equation for the phases $\theta_{n}$, which can
be written as:

\begin{equation}
\begin{aligned}\frac{d\theta_{n}}{dt}= & ak_{n+1}\frac{U_{n+1}U_{n+2}}{U_{n}}\cos\varphi_{n+1}+bk_{n}\frac{U_{n-1}U_{n+1}}{U_{n}}\cos\varphi_{n}\\
 & +ck_{n-1}\frac{U_{n-1}U_{n-2}}{U_{n}}\cos\varphi_{n-1}\\
 & +Im\left(\frac{f_{n}}{U_{n}}e^{-i\theta_{n}}\right).
\end{aligned}
\label{eq:phgoy}
\end{equation}

Note that the triadic phases defined as $\varphi_{n}=\theta_{n-1}+\theta_{n}+\theta_{n+1},$
govern both the amplitude and the phase evolution, as they determine
in particular the signs of each nonlinear term in Eqs. (\ref{eq:amp})
and (\ref{eq:phgoy}). This means that the evolution equation are
invariant under the period three transformation

\[
\begin{cases}
\theta_{3n-1} & \rightarrow\theta_{3n-1}-\delta\\
\theta_{3n} & \rightarrow\theta_{3n}+2\delta\\
\theta_{3n+1} & \rightarrow\theta_{3n+1}-\delta
\end{cases}
\]
which leaves the triadic phases invariant for any constant $\delta.$
The consequence of this $U(1)$ symmetry is that single shell phases
$\theta_{n}$ are not uniquely determined by the triadic phases, and
as also numerically confirmed, they remain uniformly distributed (i.e.
``random''). Thus, the appropriate quantities, whose dynamics and
statistics are physically relevant are the triadic phases $\varphi_{n}$,
rather than single shell phases $\theta_{n}$.

We can write the equations for the triadic phases by adding Eq. (\ref{eq:phgoy})
for three consecutive shells, which gives:
\begin{equation}
\begin{aligned}\frac{d\varphi_{n}}{dt} & =a\Lambda_{n+1}^{n}\cos\varphi_{n+1}+b\Lambda_{n}^{n}\cos\varphi_{n}+c\Lambda_{n-1}^{n}\cos\varphi_{n-1}\\
 & +a\Lambda_{n}^{n-1}\cos\varphi_{n}+b\Lambda_{n-1}^{n-1}\cos\varphi_{n-1}+c\Lambda_{n-2}^{n-1}\cos\varphi_{n-2}\\
 & +a\Lambda_{n+2}^{n+1}\cos\varphi_{n+2}+b\Lambda_{n+1}^{n+1}\cos\varphi_{n+1}+c\Lambda_{n}^{n+1}\cos\varphi_{n}
\end{aligned}
\label{eq:trph}
\end{equation}
where
\[
\Lambda_{\ell}^{n}\equiv k_{\ell}\frac{U_{\ell-1}U_{\ell}U_{\ell+1}}{U_{n}^{2}}.
\]
Note that Eq. (\ref{eq:trph}) involves the triadic phases of five
consecutive triads from $n-2$ to $n+2$, which in turn include contributions
from seven consecutive shells from $n+3$ to $n-3$, making this coupling
\emph{longer range} compared to the shell amplitude equation, which
involves only the neighboring triadic phases (i.e. $\varphi_{j}$
with $j=n-1,n,n-1$) and contributions from shells from $n-2$ to
$n+2$. Note that these couplings can also be interpreted (and generalized
to more complex interactions) within the network representation of
the shell models, see e.g. Ref. \citealp{gurcan:21}, where the evolution
of a given triad can be computed as a sum over all the triads connected
through shared nodes in the network topology.

\subsection{The Role of Triadic Phase on Energy Flux\protect\label{subsec:Choice-of-phase-1}}

The phase dynamics of a shell model with $N$ shells can therefore
be considered as being equivalent to that of a chain of $N-2$ nonlinear
oscillators, each representing a triad, where the coupling terms are
determined by a combination of the $N$ self-consistently evolving
amplitudes and the interaction coefficients. When forcing is applied
near the first few shells and dissipation occurs at high-$n$, the
chain of oscillators is constrained to accommodate the energy flux
from low to high $n$. In particular if the phases are initially distributed
in such a way that there is no initial net flux, the amplitudes will
grow until the phases will reorganize themselves in order to accommodate
the spectral flux implied by forcing and dissipation, basically because
of the \emph{free energy} associated with such a non-uniform energy
distribution. This means that the energy injection at low-$n$ drives
the phases to dynamically self-organize in order to sustain the cascade
mechanism, possibly reaching an eventual steady state.

In order to illustrate this steady state, let us consider the form
of the flux in a shell model. The energy cascades to small scales
due to nonlinear interactions, and the energy transfer in a given
triad centered at shell $n$ is defined as a local spectral energy
flux, expressed in terms of the three shell correlation $\chi_{n}=U_{n-1}U_{n}U_{n+1}e^{-i\varphi_{n}}$
as:

\begin{equation}
\begin{array}{cc}
\Pi_{n}^{E}= & 2\text{Im}\left(ak_{n+1}\chi_{n+1}-ck_{n}\chi_{n}\right).\end{array}\label{eq:goy-flux-definition-1-1-1}
\end{equation}
Using $a=\left(1+g\right)$ and $c=-\left(1+g^{-1}\right)$ and the
power law solution $\left|\chi_{n}\right|\propto k_{n}^{-1}$ , we
get 
\[
\Pi_{n}^{E}=2\left(g^{-1}+1\right)\left(g\sin\varphi_{n+1}+\sin\varphi_{n}\right)\;\text{,}
\]
which is maximized for $\varphi_{n}=\varphi_{n+1}=\pi/2$. This means
that the Kolmogorov cascade with the correspondingly power law spectrum
and a constant energy flux is maximized if the triads are aligned
at $\varphi_{n}=\varphi_{n+1}=\pi/2$. Note that since $c$ is negative,
this choice maximizes the forward flux regardless of the choice of
$\left|\chi_{n}\right|$ as well. This allows the system to deviate
locally from the Kolmogorov solution. Similarly an inverse cascade
is implied if $\varphi_{n}=\varphi_{n+1}=-\pi/2$ regardless of the
choice of the power law.

This means that the particular arrangement of the nonlinear dynamics
of the consecutive triads is important in determining both the direction
and magnitude of energy transfer through the ``boundary'' between
the two triads. In the following, we try to highlight this connection,
considering different types of shell models and studying the role
of triadic phases, and their statistical distributions in the cascade
process.

\subsubsection{Forward cascade and intermittency\protect\label{subsec:Forward-intermittent-local-1}}

In a shell model with local interactions, that conserves the inviscid
invariants typical of 3D turbulence, corresponding to Eq. (\ref{eq:goy}),
and with the standard shell spacing $g=2$, one gets a forward energy
cascade, and an energy spectrum consistent with the Kolmogorov $-5/3$
scaling together with a level of intermittency, somewhat consistent
with experiment and direct numerical simulations \citep{ohkitani:89,kadanoff1995,lvov:98}.

The multifractal corrections, implied by this intermittent behavior
(see, e.g., Refs. \citealp{ohkitani:89,kadanoff1995,pisarenko:93,de-wit:2024:}),
violating Kolmogorov's original self-similarity hypotesis, manifest
themselves as non-gaussian and scale-dependent distributions of velocity
fluctuations in the inertial range, with fatter tails at smaller scales.
For shell models, this is undoubtedly a consequence of the time evolution,
since these models lack any information about spatial distribution
of singularities. Note that, rare events in the time evolution of
shell models are linked to sudden bursts of flux that goes through
the chain of local triads that constitute them. Those events require
both large amplitudes and some kind of local phase synchronization
so that the burst can propagate across the chain in the correct direction.

More generally, the degree of locality of the cascade processes in
shell models, can be investigated through multi-shell correlation
functions \citep{biferale1999}, or via the statistical properties
of the Kolmogorov multipliers, defined as $w_{n}=\frac{U_{n}}{U_{n-1}}$,
and the triadic phases $\varphi_{n}$, as done in Refs. \citealp{benzi:93,biferale:2017,vladimirova:2021},
investigating whether or not the probability distribution functions
of those quantities are universal in the inertial range ( i.e. independent
of the shell index). It is also worth noting that the correlation
of the Kolmogorov multipliers of two different shells decays faster
than that of the triadic phases \citep{biferale:2017}. This is significant,
because it suggests that the \emph{longer range} interactions implied
by the triadic phase equation {[}i.e., Eq. (\ref{eq:trph}){]} introduce
nontrivial phase correlations that are particularly relevant for the
statistics of extreme events causing intermittency.

\begin{figure}
\includegraphics[width=0.95\columnwidth]{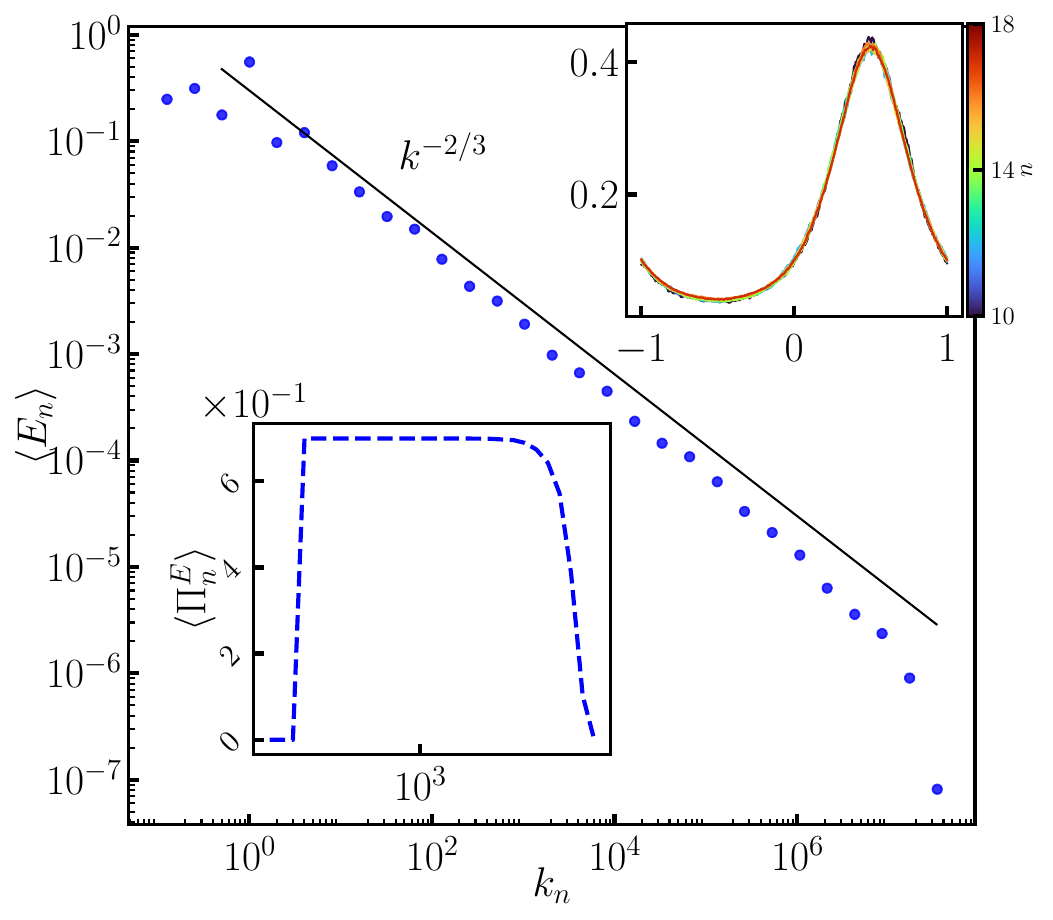}

\caption{\protect\label{fig:goy-direct-E-PI-PHI-1}Plot of the average energy
spectrum as a function of the wave number obtained for the 3D GOY
model with inter-shell spacing $g=2$, driven by large-scale forcing.
The bottom-left inset shows the average spectral energy flux as a
function of the shell wave number $k_{n}$. In the top-right inset
the normalized PDFs of the triad phases $\varphi_{n}$ are shown for
different shells corresponding to the inertial range. The shell number
is represented by different colors labeled by the continuous colorbar.}
\end{figure}

In Figure \ref{fig:goy-direct-E-PI-PHI-1}, we present the results
for a standard GOY model of three dimensional Navier-Stokes turbulence,
forced at large scales, or small $n$, which is extensively studied
in the past \citep{ohkitani:89,kadanoff1995,pisarenko:93}. The dynamics
of the model exhibit a forward energy cascade with clear inertial
range, with a spectral slope $\left|u_{n}\right|^{2}\sim k_{n}^{-2/3}$,
which is the shell model equivalent of $E(k)\propto k^{-5/3}$. In
the inset plot, the normalized PDFs of the triad phases $\varphi_{n}$
are displayed for different shells in the inertial range, showing
that they collapse to a more or less scale invariant universal distribution
similarly to Kolmogorov multipliers \citep{biferale:2017}.

This distribution exhibits a distinct peak at the fixed point $+\pi/2$
consistent with a direct energy cascade. As we will discuss later
in Sec. \ref{sec:Intermittency-1}, reducing the inter-shell spacing
progressively disrupts scale invariance and enhances intermittency
corrections.

\subsection{Sync and Kuramoto Order parameter\protect\label{subsec:Sync-and-Kuramoto-1}}

As discussed in Section \ref{subsec:Basic-Phase-Dynamics}, the cascade
in a shell model happens via self-organization of the triadic phases
$\varphi_{n}$ of the consecutive triads that constitute it, in order
to accommodate the flux that goes through them. For the forward cascade,
this causes the triadic phases to cluster around $\pi/2$, and for
the standard GOY model, this implies a scale-invariant universal distribution
of their statistics in the inertial range. In the usual scenario,
such statistics underlies intermittent dynamics, corresponding to
rare but extreme events.

In order to quantify the level of coherence of the chain of triads
that facilitates the energy cascade, we can define a global order
parameter by considering the population of nonlinear oscillators represented
by triadic phases, in order to explore the role of phase aligned triads
in extreme flux events.

For a 1D system of interacting nonlinear oscillators, which are described
by triadic phases, we can define the Kuramoto order parameter \citep{strogatz:01,strogatz:92,kuramoto:book:1984}
as:

\begin{equation}
Re^{i\Phi}=\frac{1}{N_{m}}\sum_{m}e^{i\varphi_{m}}\;\text{,}\label{eq:def_kura_glob}
\end{equation}
where the sum $\sum_{m}$ is over all the triads in the inertial range,
where a clear power low scaling is observed, independent of forcing
and dissipation.

\begin{figure}
\includegraphics[width=0.95\columnwidth]{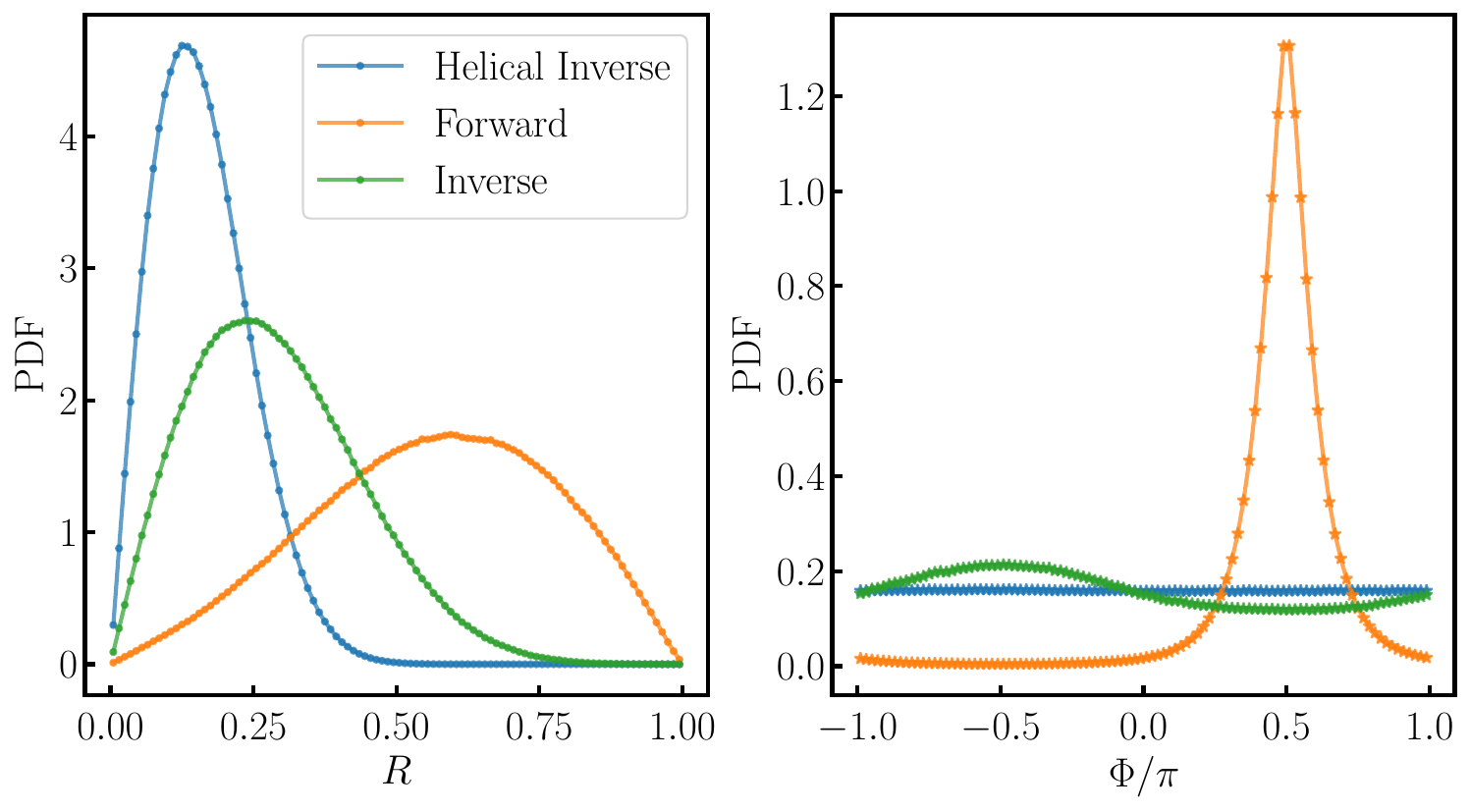}\caption{\protect\label{fig:kura-cascades}Normalized probability distribution
functions of the Kuramoto order parameter $R$ and $\Phi$, for various
shell models exhibiting different type of cascades: a classical GOY
model displaying a forward energy cascade; a helical shell model with
elongated triads driving an inverse cascade (see Sec. \ref{sec:Helical-shell-models}
and \ref{subsec:Inverse-helical} for a detailed discussion and the
definition of the Kuramoto parameter in the context of helical shell
models); and the model investigated in Ref. \citealp{tom:2017} that
drives an inverse cascade via local triads (see \ref{subsec:inverse-ray-model}
below).}
\end{figure}

At a given time $t$, the complex phase of the Kuramoto parameter
$\Phi\left(t\right)\in[0,2\pi)$ measures the \emph{average phase}
of the population of triads that are considered, and the amplitude
$R\left(t\right)\in[0,1]$ quantify the phase coherence of the chain.
A completely incoherent state corresponds to $R(t)\approx0$, since
all the triads are uniformly distributed on the unit circle, whereas,
the case with all the triad phases being perfectly aligned corresponds
to $R=1$. Using the Kuramoto parameter for a set of triadic phases,
was initially introduced in the context of the Burgers equation \citep{buzzicotti-phase:2016,murray:2018,leiva:2022}
in order to analyze the energy flux enhancement due to highly coherent
states, related to collisions of shocks in physical space. Here, we
apply it to shell models, for characterizing the dynamics of triadic
phases in driving the underlying cascade and the formation of rare
events that cause intermittency.

In Figure \ref{fig:kura-cascades} we present the normalized PDF of
both $R$, and $\Phi$ for the standard GOY model that drives a forward
cascade, discussed above and for two other models that are capable
of generating inverse cascade (which will be discussed in more detail
in Sec. \ref{sec:Inverse-cascades}, since inverse cascade in shell
models is a complicated issue). Note that the PDF of $R$ for the
forward cascade peaks at around $R\approx0.7$, it exhibits significantly
higher level of coherence with respect to the two inverse cascade
models, which peak at $R\approx0.1$ and at $R\approx0.25$ respectively.
Regarding the normalized PDF of $\Phi$, a strong evidence of phase
clustering is present for the forward cascade shell model showing
a sharp distribution with a maxima at $\pi/2$, in contrast to the
two models that result in inverse cascade displaying a broader, less
pronounced distributions. Nevertheless, the two non-intermittent models
display a weak clustering of the triadic phases around $-\pi/2$.
However, this is so weak that it is not clearly visible for the helical
inverse cascade model in the presented plot, as compared to the model
of \citet{tom:2017}.

\subsubsection{Local synchronization and energy transfer}

The clustering of triadic phases around $\pi/2$, and a significant
value of the Kuramoto parameter, as in the forward cascade case shown
in Fig. \ref{fig:kura-cascades}, provides an indicator of the global
phase coherence in the entire inertial range. We can also analyze
the local degree of synchronization within the triad chain and its
correlation to energy flux at the corresponding shell $n$, to better
understand the contribution of phase alignments in the cascade mechanism
and its variations.

In order to establish the connection between coherent structures traveling
across shells, extreme events, end the alignment or clustering of
triadic phases through the inertial range, we propose the following
local Kuramoto order parameter centered at the $n-$th shell:
\begin{equation}
R_{n}e^{i\Phi_{n}}=\frac{\sum_{t\in\Gamma(n)}w_{t}e^{i\varphi_{t}}}{\sum_{t\in\Gamma(n)}w_{t}}\;\text{,}\label{eq:kurloc}
\end{equation}
where the summation $\sum_{t\in\Gamma(n)}w_{t}$ represents a shell-dependent
weighted sum over a set of triads $\Gamma(n)$. A consistent choice,
in order to quantify the local coherence is to restrict the summation
to the set of interacting triads $\Gamma(n)$ which are connected
to the central triad of the $n$-th shell. In this approach the weight
of each triadic phase is equal to the number of times it appears on
the right-hand side of the evolution equation for $\varphi_{n}$ {[}i.e.,
Eq. (\ref{eq:trph}){]}. While the weights of those who do not appear
are set to zero. With this definition, a local coherent state ($R_{n}=1$),
corresponds to local phase locking, which implies a fixed point of
the triadic phase evolution equation. In the case of a local shell
model, this definition corresponds to a sum over a list of 5 consecutive
triads with the weights given by $\left\{ w_{n-2},w_{n-1},w_{n},w_{n+1},w_{n+2}\right\} \equiv\left\{ 1,2,3,2,1\right\} $.

\begin{figure}
\includegraphics[width=1\columnwidth]{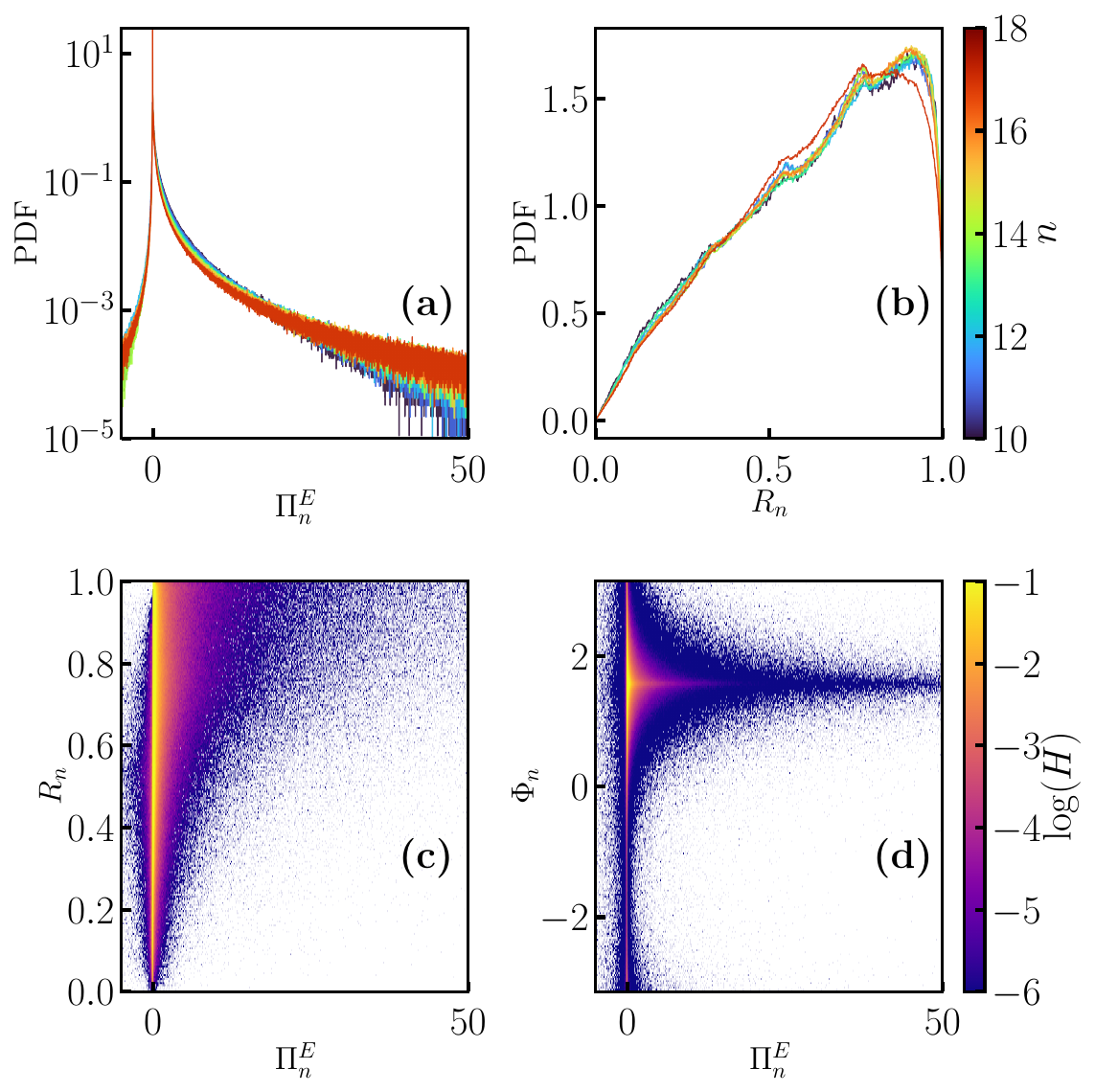}\caption{\protect\label{fig:joint_pdfs-g2}Individual and joint statistics
of energy flux, and the amplitude and phase of the Kuramoto parameter.
In panel (a), located at the top-left, the PDF of the local spectral
energy flux $\Pi_{n}^{E}$ is shown, exhibiting the typical behavior
where the distribution develops increasingly heavy tails for larger
shell indices.\textit{ }In panel (b) at the top right, the PDF of
the local Kuramoto parameter $R_{n}$ is presented, displaying a scale-invariant
universal statistics with higher probability of coherent local $R_{n}$.
Curve colors correspond to different shell numbers, as indicated by
the continuous colorbar. In panel (c) at the bottom left, the joint
PDF between $\Pi_{n}^{E}$\LyXZeroWidthSpace{} and $R_{n}$ is shown,
which illustrates how extreme flux events are predominantly correlated
with higher local synchronization events. In panel (d) at the bottom
right, the joint PDF between $\Pi_{n}^{E}$\LyXZeroWidthSpace{} and
$\Phi_{n}$ is shown\LyXZeroWidthSpace , complementary to panel (c),
that present the clustering of triad phases around $\pi/2$ which
is associated with larger flux values.}
\end{figure}

As shown in Fig. \ref{fig:joint_pdfs-g2}, panel (a), the PDFs of
the spectral energy flux $\Pi_{n}^{E}$, computed for different shells
in the inertial range, display a typical form characterized by a pronounced
peak, corresponding to a relatively small flux value (with a mean
value around $0.7$ as can be seen in the inset of Figure \ref{fig:goy-direct-E-PI-PHI-1}),
accompanied by heavy tails accounting for intense forward energy transfer
events, which can reach values as large as 500 times the mean flux.
This means that there are rare events that drive sudden and large
energy transfer.

The panel (b) of the same figure, shows the PDFs of the local synchronization
parameter $R_{n}$, computed using Eq. (\ref{eq:kurloc}), for different
shells. Notably, the distributions are scale-independent in the inertial
range and imply a clear tendency of the dynamics to favor locally
phase-coherent configurations over incoherent ones.

In panel (c), we consider the joint PDF of these two quantities for
a given shell in the inertial range, in order to investigate if local
synchronization correlates with energy transfer. The joint distribution
shows that while typical flux events, close to the mean value are
distributed across the whole range of $R_{n}$, rare but extreme flux
events only occur for large values of $R_{n}$, where consecutive
triads are partially-synchronized. In other words, while for small
$R_{n}$ the flux events are distributed more or less symmetrically
around the mean, an overwhelming majority of the large flux events,
tend to have large $R_{n}$ values. In the same spirit, panel (d)
shows the joint PDF between $\Phi_{n}$ and the local energy flux
$\Pi_{n}^{E}$, illustrating that the extreme events are associated
with a pronounced clustering of triad phases around $\pi/2.$ These
results show clearly that rare events go hand in hand with local synchronization
of triadic phases in shell models. In the following we will study
the effect of these rare events on a well-established measure of turbulence
intermittency and how this behavior varies with the inter-shell spacing
$g$, which changes both the \textit{synchronizability} of the triadic
interactions and the number of its elements.

\section{Intermittency\protect\label{sec:Intermittency-1}}

We have shown that extreme events in energy flux (and amplitudes $U_{n}$)
are linked to triadic phase coherence. The statistical properties
of these events determine the level of intermittency, manifesting
itself in the deviation of the scaling of the structure functions,
defined as $S_{p}(k_{n})=\left\langle \left|u_{n}\right|^{p}\right\rangle ,$
from the scaling $S_{p}\propto k_{n}^{-p/3}$ predicted by Kolmogorov's
self-similarity hypothesis. In the GOY shell model, intermittency
corrections can be more accurately quantified by studying the scaling
behavior of the three point correlations:

\[
\Sigma_{p}(k_{n})=\left\langle \left|Im\left(\chi_{n}\right)\right|^{p/3}\right\rangle \;\text{,}
\]
which eliminates period three oscillations of shell amplitudes \citep{kadanoff1995}.
The scaling exponents $\xi(p)$ of the $p$-th order structure function
can be estimated either by performing a linear fit in a log-log plot
of $S_{p}(k_{n})$ or $\Sigma_{p}(k_{n})$ versus $k_{n}$, or trough
a more refined method such as the Extended Self-Similarity (ESS) \citep{benzi93}.

The ESS approach, estimates the scaling exponents $\xi_{ESS}(p),$
as the power low scaling of the structure function of order $p$,
with respect to the third order one, i.e. the scaling $S_{p}(k_{n})\sim S_{3}^{-\xi_{ESS}(p)}$
($\Sigma_{p}(k_{n})\sim\Sigma_{3}^{-\xi_{ESS}(p)}$). This approach
can reveal hidden scalings in turbulent structure functions (see Ref.
\citealp{banerjee:2013} as an example in the shell model context).
The improved accuracy of the ESS method has been explained as its
ability to suppress subdominant corrections \citep{chakraborty:2010}
and an interpretation of ESS within the multifractal formalism has
been proposed in Ref. \citealp{benzi:1996}.

In the following analysis, we present different scaling exponents
$\xi(p)$ that are extracted from numerical simulations of different
shell models. In particular, for the GOY model we employ the three-point
correlation $\Sigma_{p}(k_{n})$, using the ESS procedure, restricting
the fits to a range of scales corresponding to the constant flux window
in the inertial range. Note that even though no hidden scaling regime
is revealed by the ESS analysis (i.e. $\xi_{ESS}(p)\simeq\xi(p)$);
here, it helps refine our estimation of the intermittency.

\begin{figure}
\includegraphics[width=1\columnwidth]{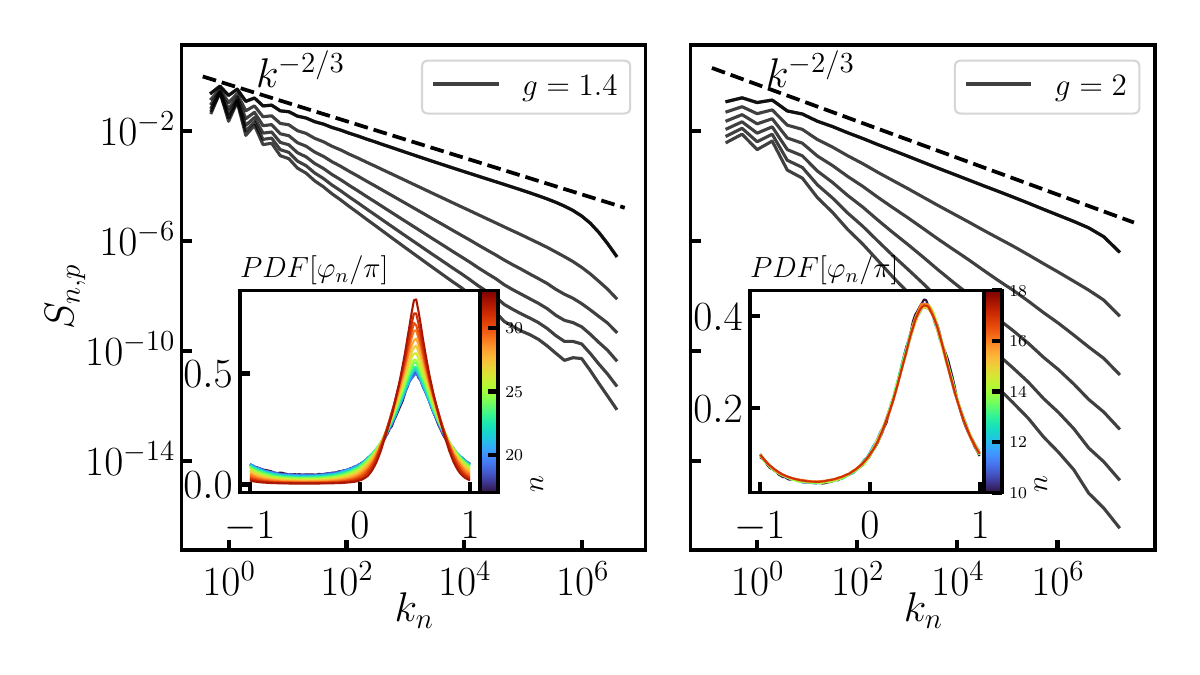}

\caption{\protect\label{fig:scaling-S_p=000026-pdf-phi}Scaling of the structure
functions $S_{p}(k_{n})$ (orders $p=2,3,\dots6,7$) as a function
of the shell wave-number $k_{n}$, shown for two representative values
of the inter-shell spacing $g=1.4$ and $g=2$. The right panel ($g=2$)
presents the typical intermittency corrections, while the left panel
($g=1.4$) presents an enhancement of intermittency. The insets display
the PDFs of the triadic phases $\varphi_{n}$ for different shells
in the inertial range, with colors indicating shell index $n$ as
shown in the colorbar.}
\end{figure}

\subsection{Shell Spacing Dependence\protect\label{subsec:Shell-spacing-Dependence}}

Although the widely adopted choice in literature, is to set the inter-shell
spacing $g=2,$ having actual triadic interactions involving a Fourier
grid structure with the wave-numbers given by $k_{n}$ requires an
inter-shell ratio smaller than the golden ratio. It is therefore interesting
to study how the intermittency corrections depend on the inter-shell
spacing $g$. In a recent work \citep{manfredini:2025}, exploring
a class of asymptotically self-similar shell models based on recurrent
sequences -where the shell wave numbers $k_{n}$ are integers- we
observed that the intermittency corrections increase for sequences
characterized by smaller inter-shell spacings.

\begin{figure}
\includegraphics[width=0.7\columnwidth]{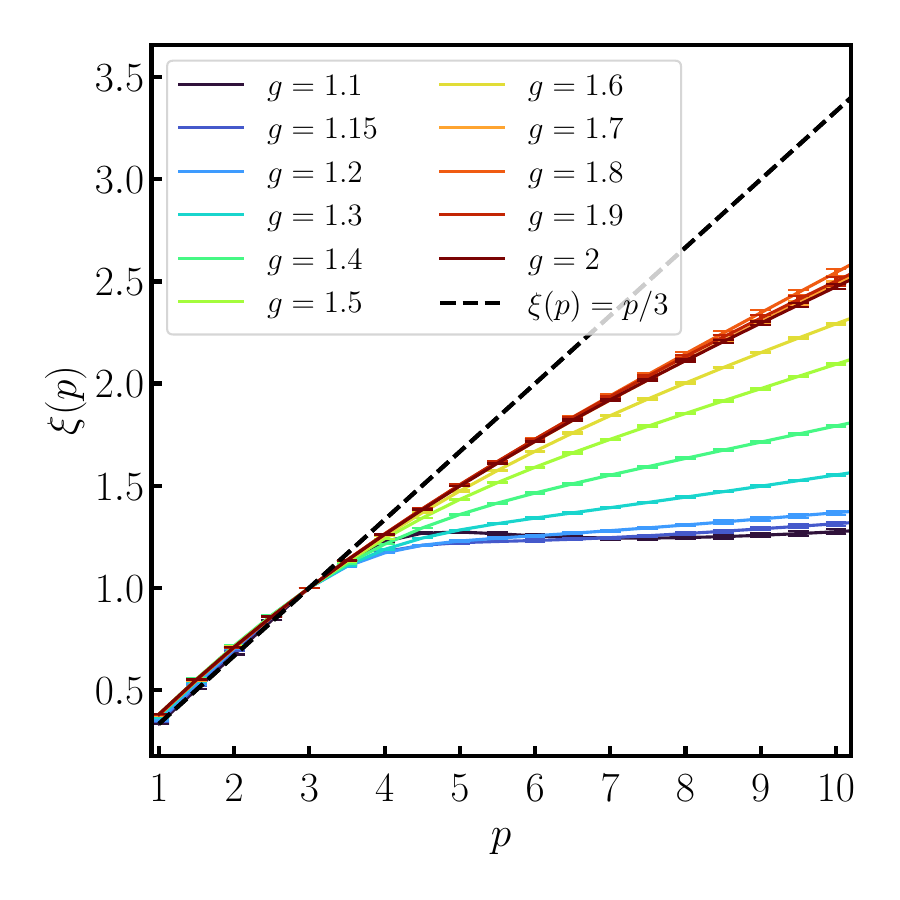}

\caption{\protect\label{fig:xip-vs-p}The scaling exponent $\xi(p)$ of the
$p-$th order structure function as a function of $p$, for different
shell spacings. The dashed line represents the Kolmogorov $p/3$ scaling.}
\end{figure}

\begin{figure}
\includegraphics[width=0.7\columnwidth]{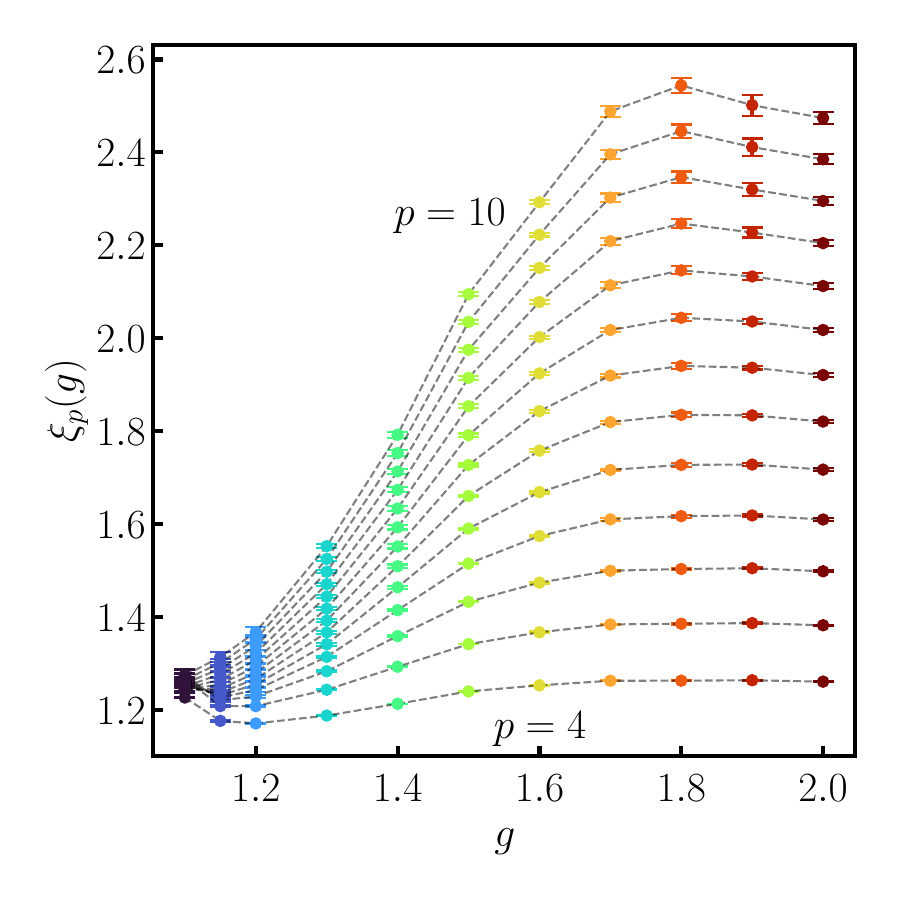}

\caption{\protect\label{fig:xip-vs-g}Scaling exponent $\xi_{p}(g)$ for structure
functions of order $p$ (ranging from 4 to $10$ in steps of $0.5$),
plotted as a function of the inter-shell spacing $g$. The marker
colors correspond to different shell spacings, as in Fig. \ref{fig:xip-vs-p}.}
\end{figure}

In order to study this phenomenon, here we provide a detailed characterization
of the intermittency as a function of inter-shell spacing in regular
shell models, in particular considering the role of triadic phase
synchronization in the generation of rare and intense events of energy
transfer that are responsible for the increased intermittency.

We first note that the intermittency corrections have a non-trivial
dependence to the inter-shell spacing $g$ for the standard (GOY)
shell model. Figure \ref{fig:xip-vs-p} shows the estimated scaling
exponents $\xi(p,g)$ of the $p$-th order structure functions, for
different values of the spacing, in the range $g=1.1-2$, suggesting
a progressive enhancement of intermittency, corresponding to a flattening
of the scaling exponent curves as $g$ is decreased. These observations
are consistent with previous studies, particularly those investigating
the continuum limit of shell models, where $g\to1$ leads to equations
closely resembling the Burgers equation \citep{Andersen:00}, known
to generate shocks and an intermittent, bifractal anomalous scaling
behavior \citep{bec:2007}. Further insight is provided by Figure
\ref{fig:xip-vs-g}, which shows the dependence of the scaling exponents
$\xi_{p}(g)$ as a function of the inter-shell spacing $g$. High
order scaling exponents such as $p\gtrsim8$ exhibit a sharp transition
when $g$ decreases below a critical value $g_{c}\simeq1.7$, suggesting
a qualitative change in the scaling properties of the system. It seems
that for $g\apprge g_{c}$ the scaling is roughly independent of the
inter-shell spacing, while for $g\apprle g_{c}$ the anomalous scaling
exponents vary significantly with the inter-shell spacing, following
an approximately linear scaling.

Note also how $g_{c}$ is just slightly above the golden ratio. While
this may be a pure coincidence, there may be a link to nonlinear frequencies
$\omega_{n}$, of the $n$-th triad being close enough to interact
more efficiently through a recurrence relation such as $\omega_{n}=\omega_{n-1}+\omega_{n-2}$.

\subsubsection{Shell spacing and the level of synchronization}

With the aim of highlighting the role of synchronization events in
increasing intermittency corrections, we analyze the statistical properties
of triadic phases across the inertial range. In Figure \ref{fig:scaling-S_p=000026-pdf-phi},
we present the scaling of structure functions for two representative
cases, $g=2$ and $g=1.4$, which are characterized by a standard
and very high levels of intermittency, respectively. In the inset
of the same figure, we present the PDFs of triadic phases for different
shells within the inertial range. For the higher intermittency case
($g=1.4$), the triadic phase distributions exhibit a pronounced scale-dependent
behavior, becoming progressively narrower at smaller scales in the
inertial range, a similar trend is also observed in the statistics
of the Kolmogorov multipliers (not shown in the plot). This behavior
is actually common for all the $g$ values below the critical value
$g\apprle g_{c}$ and becomes more pronounced as $g$ decreases. This
suggests a tendency towards stronger phase alignment at higher wave
numbers, consistent with the idea that the increase in intermittency
is due to rare but large dissipation events at small scales.

\begin{figure}
\includegraphics[width=0.8\columnwidth]{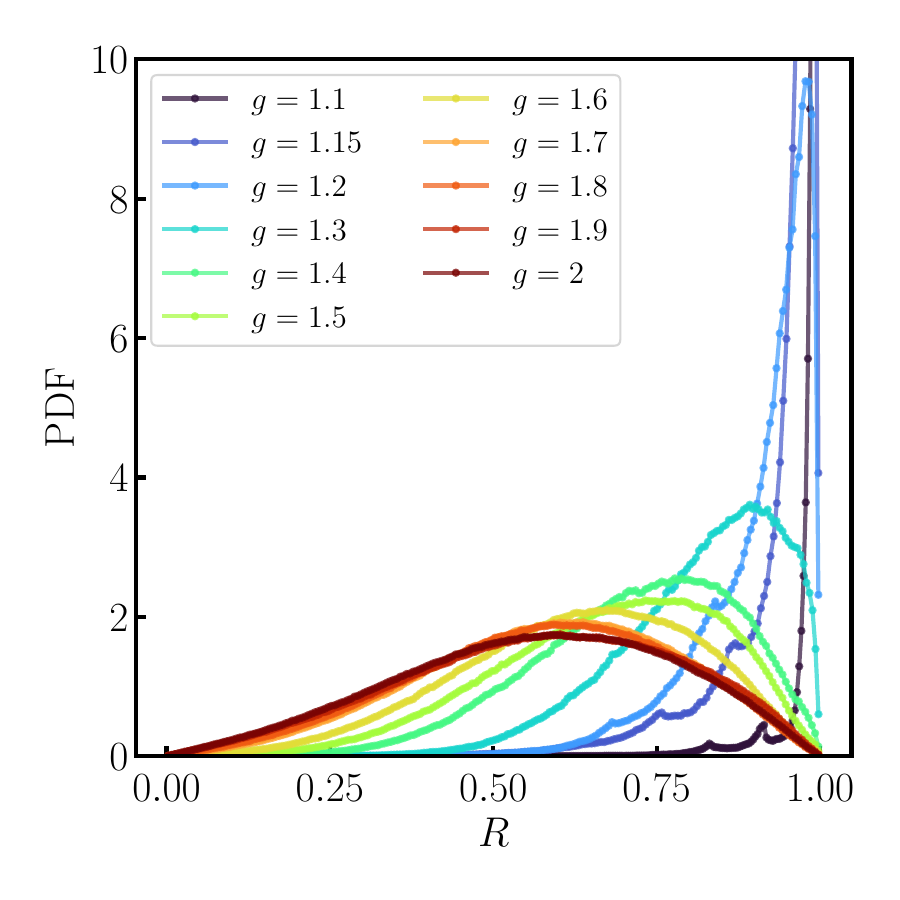}

\caption{\protect\label{fig:kuramoto-r-g-pdf}Normalized PDFs of the Kuramoto
order parameter R, shown for the different values of the inter-shell
spacings $g$ that are considered.}
\end{figure}

In order to characterize the global level of synchronization across
the inertial range, we consider the complex Kuramoto order parameter,
defined in Eq. (\ref{eq:def_kura_glob}), and present the PDFs of
the coherence parameter $R$ in Fig. \ref{fig:kuramoto-r-g-pdf} as
a function of the inter-shell spacing $g$.

We note that, as the inter-shell spacing decreases, the PDF of the
coherence parameter $R$ become increasingly skewed and a finite probability
for fully synchronized events appears. These are states in which all
the triads are aligned, and are associated with rare, extreme events
in the forward energy flux. Remarkably, we observe that below the
critical value $g\apprle g_{c}$, the PDFs of $R$ are $g$-dependent,
while for for $g\apprge g_{c}$ they collapse to the same distribution,
which is also in accordance with the behavior of intermittency corrections
as a function of $g$. This suggests that the Kuramoto parameter,
despite being only a measure of the phase coherence, can be used as
a proxy for detecting different levels of intermittency.

\begin{figure}
\includegraphics[width=0.8\columnwidth]{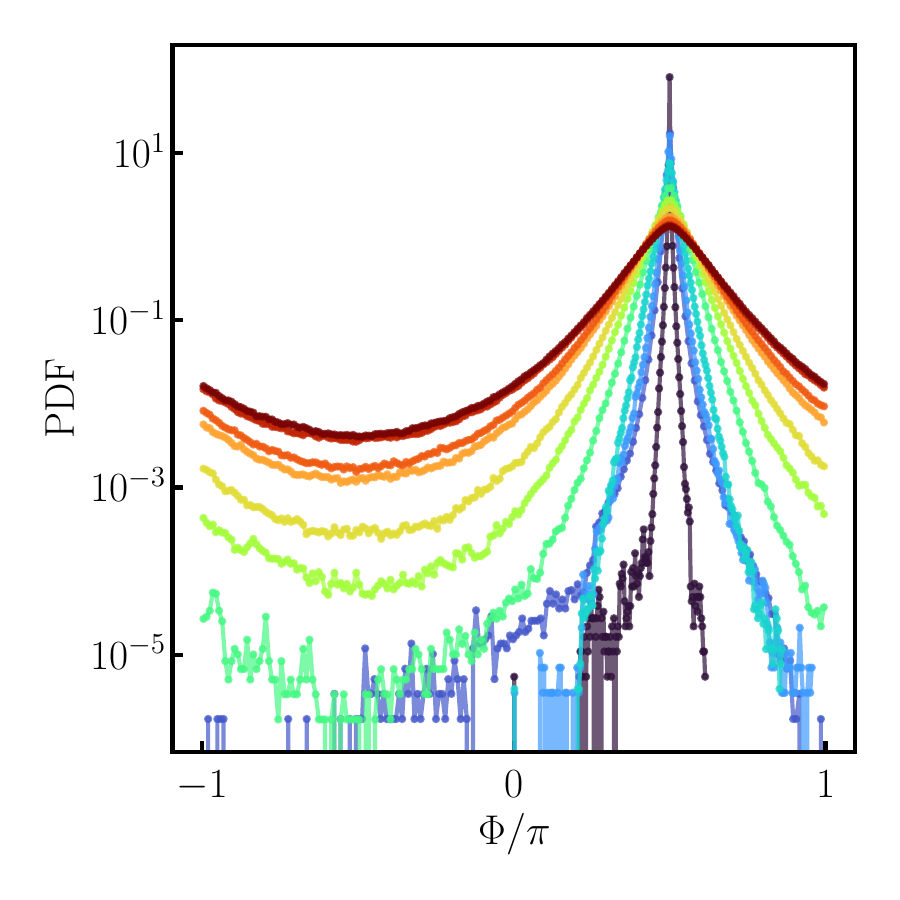}\caption{\protect\label{fig:kuramoto-phi-g-pdf}Normalized PDFs of the average
phase $\Phi$, for different shell spacings $g$. Marker colors correspond
to the different shell spacings, as in Fig. \ref{fig:kuramoto-r-g-pdf}.}
\end{figure}

The statistics of the average phase, represented by the phase $\Phi$
of the Kuramoto parameter is presented in Fig. \ref{fig:kuramoto-phi-g-pdf}.
The PDFs of $\Phi$ as a function of $g$ are all centered around
$\pi/2$, but with a width that decreases rapidly for $g\apprle g_{c}$
, indicating that for smaller $g$ the system has a stronger tendency
to cluster around a common phase that maximizes the forward energy
flux.

\begin{figure}
\includegraphics[width=1\columnwidth]{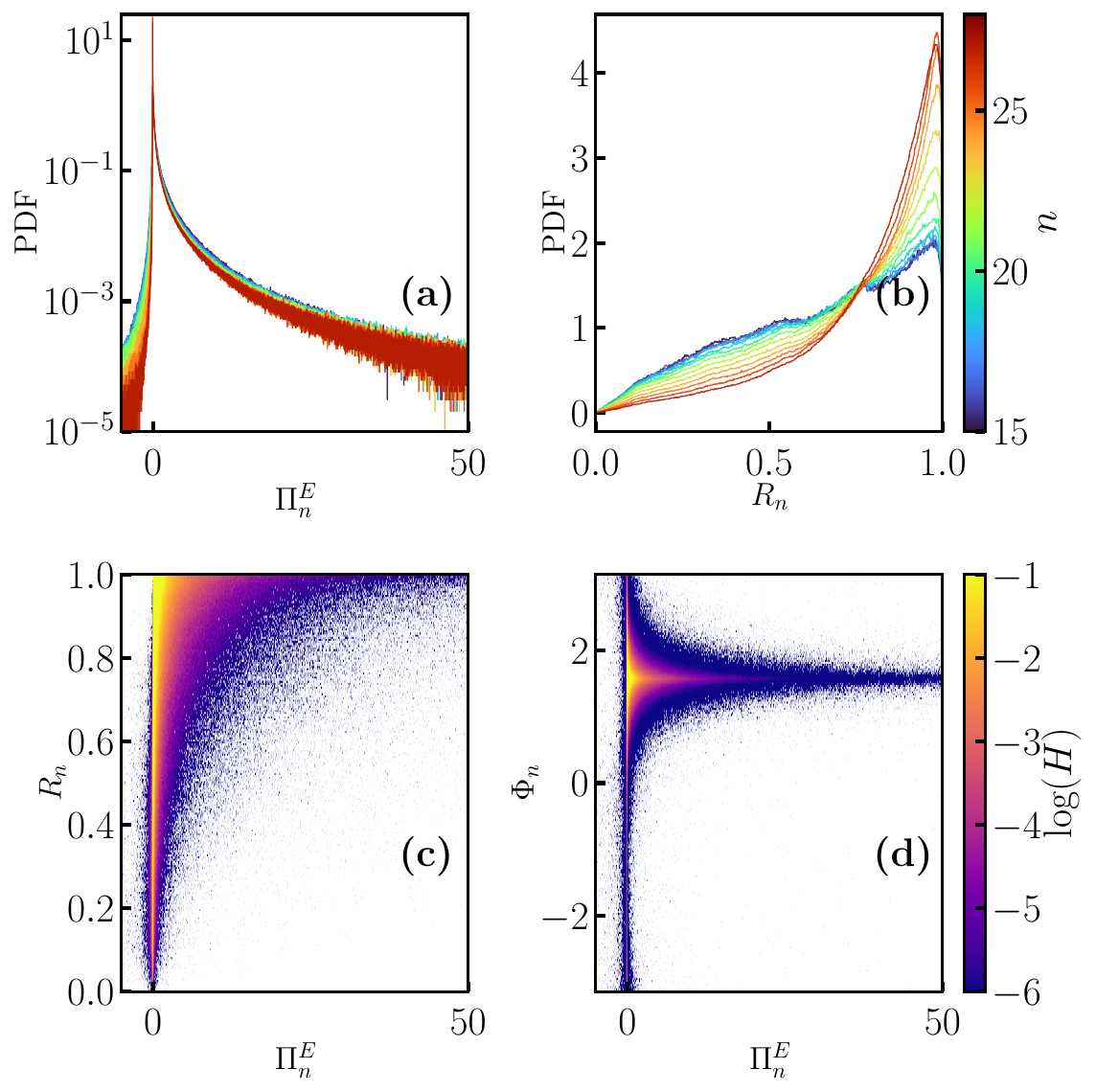}

\caption{\protect\label{fig:joint_pdfs-g15}Individual and joint statistics
of energy flux and the amplitude and phase of the Kuramoto parameter.
In panel (a) on the top left plot, the PDF of the local spectral energy
flux $\Pi_{n}^{E}$ is shown, exhibiting the typical behavior where
the distributions becomes progressively more heavy-tailed for shells
corresponding to smaller scales. This reflects the increasing occurrence
and intensity of energy bursts as we move towards smaller scales in
the inertial range.\textit{ }In panel (b) on the top right, the PDF
of the local Kuramoto phase coherence $R_{n}$ is presented, where
smaller scales display higher synchronization levels, with an increasing
probability of local fully coherent states ($R_{n}\simeq1$). Curve
colors correspond to different shell numbers, as indicated by the
continuous colorbar. In panel (c) on the bottom left, the joint PDF
between $\Pi_{n}^{E}$\LyXZeroWidthSpace{} and $R_{n}$ is shown, which
illustrates how extreme flux events are predominantly associated with
high local synchronization events. In panel (d) on the bottom right,
the joint PDf between $\Pi_{n}^{E}$ and $\Phi_{n}$ is shown, complementary
to panel (c), reinforcing the observation that extreme flux events
are associated with a marked clustering of triad phases around $\pi/2$.
Note that this state, where the phases are locked, is highly efficient
in transferring energy towards small scales.}
\end{figure}

Not surprisingly, global synchronization goes hand in hand with local
coherence. This can be seen clearly in Figure \ref{fig:joint_pdfs-g15},
which is equivalent to Fig. \ref{fig:joint_pdfs-g2}, but with an
inter-shell spacing of $g=1.5,$ corresponding to strong intermittency.
In Panel $(a)$, the statistics of the local spectral energy flux
$\Pi_{n}^{E}$ are presented, showing an increasing occurrence and
intensity of energy transfer along with a reduced probability of negative
flux events as we move towards smaller scales. This behavior is evidenced
by the progressively heavier tails of the distribution curves. The
scale-dependent organization of triadic phases $\varphi_{n}$ can
be seen in the statistics of the local phase coherence parameter $R_{n}$,
in panel $(b)$, which show that smaller scales are associated with
higher synchronization levels, and an increased probability of fully
coherent states ($R_{n}\simeq1$). These events are clusters of locally
aligned triads with $\varphi_{n}\simeq\pi/2$ corresponding to a fixed
point of the phase dynamics.

The correlation between the energy transfer and the local synchronization
level along the triad chain is analyzed via the joint statistics of
$\Pi_{n}^{E}$ and $R_{n}$, as well as $\Pi_{n}^{E}$ and $\Phi_{n}$,
shown in panel (c) and (d) respectively. These plots demonstrate that
local synchronized triad configurations - identified by $R_{n}\simeq1$
and $\Phi_{n}\simeq\pi/2$ - are strongly correlated with extreme
flux events.

\subsection{Burst events and temporal intermittency\protect\label{subsec:Burst-events-and}}

So far, we have characterized the statistical properties of shell
models as a function of the inter-shell spacing $g$, showing the
relation between phase coherence and statistics of extreme flux events,
leading to different levels of intermittency.

In particular, we showed that the flattening of the scaling exponents
$\xi(p)$ of the structure functions that is observed for decreasing
values of $g$, is rather similar to what happens in Burgers equation
with the formation of shocks. The equivalent process in shell models
is the formation of burst-like solutions. These solutions have been
addressed in the continuum limit analytically by assuming a zero background
by \citet{Andersen:00}.

Similar burst events are also present in the fully developed turbulent
state, coexisting with a background power law solution of the average
spectral energy. They can be identified as coherent sequences of local
maxima in the time evolution of the shell amplitudes $U_{n}$, propagating
across shells. Such structures, sometimes referred to as \emph{instantons},
and have been proposed as a key mechanism underlying anomalous scaling
in shell models \citep{mailybaev:12,mailybaev:2013}.

A detailed characterization of the statistical properties of instantons
as a function of the inter-shell spacing $g$ lies beyond the scope
of the present work. This would represent an interesting future direction,
particularly in the low-$g$ regime, characterized by stronger intermittency,
where we observe also backward-propagation events, suggesting a scenario
along the lines of a gas of colliding structures, rather than isolated,
unidirectional bursts.

Nevertheless, continuing with the line of inquiry of this work, we
present a phenomenological analysis of these burst like events with
respect to the local synchronization properties of the system.

\begin{figure}
\includegraphics[width=0.95\columnwidth]{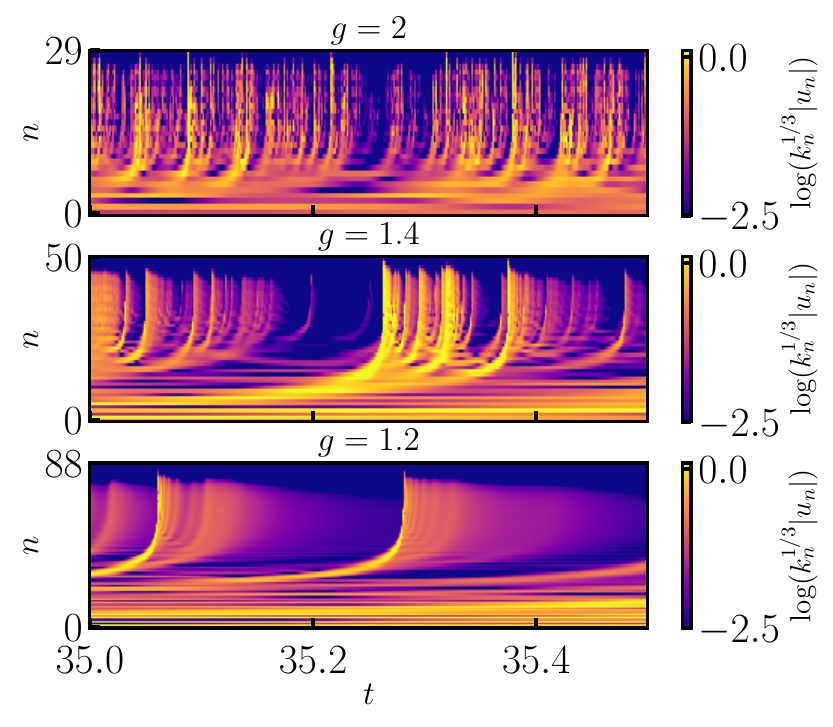}

\caption{\protect\label{fig:burstcomp-1} Temporal evolution of the compensated
shell amplitudes, $log(k_{n}^{1/3}U_{n}$) along the one dimensional
chain (shell index $n$ in the $y$-axis) shown for different inter-shell
spacing $g=2,1.4,1.2$. (top to bottom panels). The plots illustrate
how the coherent events became more and more relevant as $g$ decreases.}
\end{figure}

First we present how the dynamics of these events depend on the degree
of intermittency, controlled by the inter-shell spacing $g$. In order
to visualize these structures, Fig. \ref{fig:burstcomp-1} provides
the temporal evolution of the compensated shell amplitudes, for different
values of the inter-shell ratio $g=2,1.4,1.2$. We can see rather
frequent but less intense bursts for larger $g$, some of which may
dissipate or annihilate within the inertial range, before reaching
the smallest scales. In contrast, as $g$ is decreased, bursts become
less frequent, more localized, and significantly more intense, allowing
us to see their propagation across the whole inertial range. In this
regime, a given shell spends most of its time in a quiescent state,
interrupted by the passage of large-scale coherent bursts, which may
also occasionally be accompanied by secondary events forming tree-like
patterns.

\begin{figure*}
\includegraphics[width=0.95\textwidth]{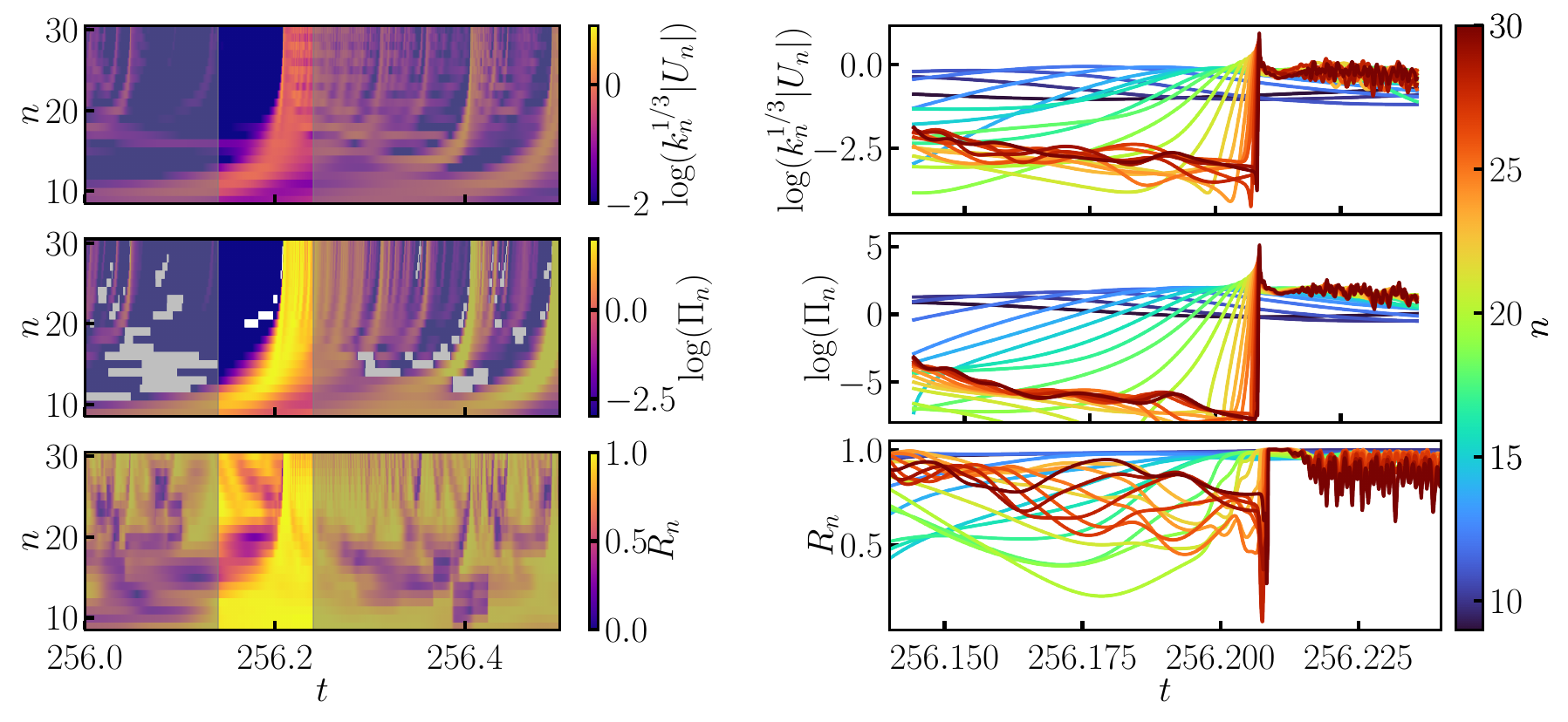}

\caption{\protect\label{fig:burst_phases-1}The left panels show colormaps
of the temporal evolution of the compensated amplitudes, $log(k_{n}^{1/3}U_{n}$)
, the energy flux, $log(\Pi_{n}^{E}$), and the local Kuramoto parameter,
$R_{n},$ across different shells (y-axis). The right panels present
the temporal evolution of the same quantities during a burst event
(unshaded region in the left panels), plotting distinct curves for
different shells (as indicated by the colorbar) highlighting the local
synchronization dynamics associated with large deviation events.}
\end{figure*}

In order to support our hypothesis that the dynamics of these bursts
are associated with triadic phase synchronization, we also present
the time evolution of the compensated amplitudes, the energy flux,
and the local Kuramoto parameter $R_{n}$ in Figure \ref{fig:burst_phases-1},
for inter-shell spacing $g=1.5$. This demonstrates that the arrival
of a burst corresponds to a rapid growth of local synchronization
level, which precedes the extreme flux event. It also shows that as
the flux event arrives at the smallest scales, all the shells are
synchronized and they loose this synchronization slowly over time
with the smallest scales becoming desynchronized first. In this sense,
synchronization starts from large scales and propagates towards small
scales while desynchronization starts from small scales and propagates
towards large scales.

One explanation of these observations is that the energy, being injected
at large scales, tends to accumulate in low$-n$ shells, until the
triadic phases can organize coherently, facilitating the release of
this accumulated energy across the shells. This naturally starts at
the large scales, and proceeds like a domino effect towards small
scales. Therefore, as the burst propagates, the local Kuramoto parameter
exhibits a sharp increase, corresponding to a local synchronization
of the system. This is necessary to accommodate the passage of the
large deviation in flux. After the passage of the burst, the system
remains partially phase locked for some time, corresponding to a cluster
of scales with $R_{n}\simeq1$, maintaining an \textquotedblleft open
channel\textquotedblright{} for energy transfer (in terms of triadic
phases). This synchronized configuration corresponds to a fixed point
of the triadic phase equation, which provides a quasi-stationary state.
It then gets desynchronized at the smallest scale first, probably
because of the nonlinear time scales associated with the smallest
scales are the shortest, so the decay to a desynchronized state in
the absence of a synchronizing force is faster at small scales. Although
we do not present the analogue of Fig. \ref{fig:burst_phases-1} for
different values of the inter-shell spacing, we note that as $g$
decreases, the phase locked state appears to be more robust and stable,
requiring a longer time for the system to desynchronize and return
to a disordered phase configuration.

After the decay of the burst, energy injected at large scales starts
to accumulate again in the low-$n$ shells, and a new burst is eventually
triggered, in a sort of cyclic dynamics of energy accumulation and
release. (Interesting our observation echos, with the results of Ref.
\citealp{benzi:2022} in which avalanche-like behavior of seismic
events, has been compared to energy release events and their inter-event
times (waiting times). In our framework, the duration of the quiescent
states --- i.e., the time intervals during which energy accumulates
at large scales --- appears to be directly influenced by the inter-shell
spacing $g$.

These observations suggest that the increased stability of phase-locked
states --- and consequently the longer duration of the so-called
quiescent periods during which energy is accumulated in low-$n$ shells
--- may favor the occurrence of extreme events capable of releasing
larger amounts of energy.

\section{Helical shell models and sync\protect\label{sec:Helical-shell-models}}

We have shown that the existence of phase coherent energy transport
events account for the anomalous scalings of structure functions and
therefore intermittency in shell models, and that the phase coherence
happens through a mechanism of nonlinear synchronization. This means
that the affinity of its triads to synchronization, is the key feature
of a shell model that determine its intermittency. So far, we have
characterized the \textit{synchronizability} of only the simplest
shell model: the classical GOY model with a one dimensional chain
consisting in $N$ shells connected by $N-2$ triads.

However, the \textit{synchronizability} of a system of triads (or,
generally, a system of coupled oscillators) also depends on the way
its elements are wired, which is represented by its topology. It is
somewhat obvious that topological effects become relevant as we depart
from a linear chain picture, as in the case of more complex model
such as nested polyhedra models or log-lattices, where multiple elements
interact at each scale, but even if we put aside those more complex
examples, a minimal reintroduction of network topology appears in
simple shell models if we try to incorporate helicity in a proper
way. This happens if we consider two signs of helicity in each scale
-instead of one sign in odd scales and another sign in even scales-
as in helical shell models. These models are a natural generalization
of the standard shell model, in which each scale is associated with
two dynamical variables, $u_{n}^{+}$ and $u_{n}^{-}$, carrying positive
and negative helicities, respectively. In this framework the two inviscid
invariants are defined as:

\begin{equation}
E=\sum_{n}|u_{n}^{+}|^{2}+|u_{n}^{-}|^{2},\quad H=\sum_{n}k_{n}(|u_{n}^{+}|^{2}-|u_{n}^{-}|^{2}).
\end{equation}

This formulation results in two, one-dimensional sequences of shell
variables, corresponding to the two helicity signs. The triadic interactions
among those modes can be systematically recovered from Waleffe's helical
decomposition \citep{waleffe:92}, where the nonlinear term in the
Navier-Stokes equations is decomposed into distinct classes of helical
interactions.

The general form of a helical shell model, omitting the dissipative
and forcing terms, with arbitrary non-local triads connecting the
three shells $n-l,n,n+m$, can be written compactly as:

\[
\frac{du_{n}^{s_{0}}}{dt}=s_{0}\left(a_{n}u_{n+\ell+m}^{s_{3}*}u_{n+\ell}^{s_{1}*}+b_{n}u_{n-\ell}^{s_{1}*}u_{n+m}^{s_{2}*}+c_{n}u_{n-m}^{s_{2}*}u_{n-\ell-m}^{s_{3}*}\right)
\]
with

\begin{align*}
 & a_{n}=Q\frac{s_{1}}{s_{0}}\left(\frac{s_{2}}{s_{0}}k_{n+\ell+m}-k_{n+\ell}\right)\\
 & b_{n}=Q\left(\frac{s_{1}}{s_{0}}k_{n-\ell}-\frac{s_{2}}{s_{0}}k_{n+m}\right)\\
 & c_{n}=Q\frac{s_{2}}{s_{0}}\left(k_{n-m}-\frac{s_{1}}{s_{0}}k_{n-\ell-m}\right)
\end{align*}
$s_{3}\equiv s_{0}s_{1}s_{2},$ and where $Q=Q_{\ell m}\left(s_{1}/s_{0},s_{2}/s_{0}\right)$
is the geometric factor of the triad class that is considered. The
four triad classes can be defined as the $4$ possible combinations
of the signs of $s_{1}/s_{0}$ and $s_{2}/s_{0}.$ Notice that the
choice of local interactions corresponds to setting $l=1,m=1$, and
results in $2N$ modes connected by $2N-4$ triads for each class.
This represents - with the usual choice $g=2$ - the helical shell
model introduced in Ref. \citealp{benzi:96}, investigated by considering
the different classes of interactions separately.

However, the general formulation of the model allows us to consider
arbitrarily elongated triads as investigated by \citet{depietro:15}.
Moreover, one can also consider a right-hand side that constitutes
a sum over different classes, each corresponding to different geometrical
factors defined in terms of the shell index and the elongation of
the triads \citep{rathmann}.

Each of these choices correspond to a different network of triadic
interactions, connecting two sequences of helical modes. For simplicity,
we focus on models constructed from a single interaction class, and
adopt the labeling SM1, SM2, etc. (see , e.g. , Refs. \citealp{benzi:96,depietro:15})
following Benzi and De Pietro. There are qualitative differences among
the networks represented by different classes. Note in particular
that for certain models, the triadic interactions form two independent
chains that do not interweave. We call those models \emph{separable.}

This allows us to distinguish two types of topologies namely \textit{separable
}and\textit{ non separable.} The SM4 class, corresponds to $s_{1}/s_{0}=$$s_{2}/s_{0}=+$,
is \textit{separable} since it consists of homochiral triads of the
type $(+,+,+)$ or $(-,-,-)$ where two helicities decouple and each
helical variable evolves independently that conserve separately their
corresponding positive-definite invariants, (i.e. $\sum_{n}k_{n}\left|u_{n}^{+}\right|^{2}$and
$\sum_{n}k_{n}\left|u_{n}^{-}\right|^{2}$), making it a candidate
to drive an inverse cascade, which is in practice overwhelmed by equipartition.
The other \textit{separable} class is the SM1 class, corresponding
to $s_{1}/s_{0}=-$, and $s_{2}/s_{0}=-,$ involving heterochiral
triads of the type $(-,+,-)$ or $(+,-,+)$, which naturally splits
into two distinct chains of modes with alternating helicities, consisting
effectively of two uncoupled GOY models.

The remaining classes SM3 and SM2, are \textit{non separable}\textit{\emph{,
since it is not possible to disentangle the triad chains in these
models into two distinct chains because of the way the interactions
are coupled together. This means that for these classes the whole
chain is interwoven, and the energy transfer across such a chain involves
multiple interactions at each scale,}} which causes the \textit{synchronizability}
to be qualitatively different.

In this case, the right-hand side of the phase evolution equation
involves contributions from up to seven different triads, in contrast
to the five involved in the \emph{separable} (single-chain) case,
as detailed in Sec. \ref{subsec:Basic-Phase-Dynamics}. Therefore,
in order to characterize the level of synchronization in a \textit{non
separable} chain, we define a weighted local Kuramoto parameter, constructed
by summing the contributions of up to seven phases of its interacting
triads.

More specifically, in a helical shell model, one can identify two
distinct triads centered at the same shell, which can be written for
the $n$-th shell as $u_{n-1}^{s_{1}}u_{n}^{s_{0}}u_{n+1}^{s_{2}}=\chi_{n}e^{i\varphi_{n}}$
and $u_{n-1}^{-s_{1}}u_{n}^{-s_{0}}u_{n+1}^{-s_{2}}=\bar{\chi}_{n}e^{i\bar{\varphi}_{n}}$.
Note that the evolution equations are odd under the parity transformation,
which implies that an organization of triad phase that maximizes the
forward energy flux corresponds to the choice $\varphi_{n}=\pi/2$,
$\bar{\varphi_{n}}=-\pi/2$. We then define the local weighted Kuramoto
parameter as:

\begin{equation}
R_{n}e^{i\Phi_{n}}\equiv\frac{\sum_{t\in\Gamma(n)}w_{t}e^{i\varphi_{t}}+\bar{w}_{t}e^{-i\bar{\varphi}_{t}}}{\sum_{t\in\Gamma(n)}w_{t}+\bar{w}_{t}}\;\text{,}\label{eq:kurloc-1}
\end{equation}
 where the weights
\[
\left\{ w_{n-2},w_{n-1},w_{n},w_{n+1},w_{n+2};\bar{w}_{n-2},\bar{w}_{n-1}\bar{w}_{n},\bar{w}_{n+1},\bar{w}_{n+2}\right\} 
\]
are determined by the structure of the phase evolution equation that
correspond to a set of interacting triads $\Gamma(n)$ which are connected
with the central triad of the $n$-th shell $u_{n-1}^{s_{1}}u_{n}^{s_{0}}u_{n+1}^{s_{2}}=\chi_{n}e^{i\varphi_{n}}.$
These weights, consistently with the previous definition (Eq. (\ref{eq:kurloc})),
are equal to the multiplicity of each triad in the right-hand side
of the evolution of equation for the triadic phase $\varphi_{n}$.
In helical shell model, their values depends on the topology of the
interaction class. For instance, for the triads of type SM3 the weights
are $\left\{ 0,1,3,1,0;1,1,0,1,1\right\} $, while for the \emph{separable}
class SM1, they are $\left\{ 1,0,3,0,1;0,2,0,2,0\right\} .$ Please
note how these weights are consistent with the definition adopted
for the single chain case, that we can express as $\left\{ w_{n-2},\bar{w}_{n-1},w_{n},\bar{w}_{n+1},w_{n+2}\right\} $,
since the SM1 reduces to two uncoupled GOY models.

In the following, we will characterize the level of synchronization
in helical shell models with respect to their cascade properties and
the presence of intermittency corrections.

\subsection{Shell spacing dependence}

Helical shell models provide a rich framework, in which a set of interacting
triads are connected in a way that mirrors the actual interactions
among two sequences of helical modes in three dimensional Navier-Stokes
turbulence. Here, we consider the phase coherence of these two sequences,
as a function of the inter-shell ratio $g$, in order to investigate
how differences in \textit{synchronizability} are related to energy
transfer across scales and to compare with the single chain GOY model.

\begin{figure}
\includegraphics[width=0.85\columnwidth]{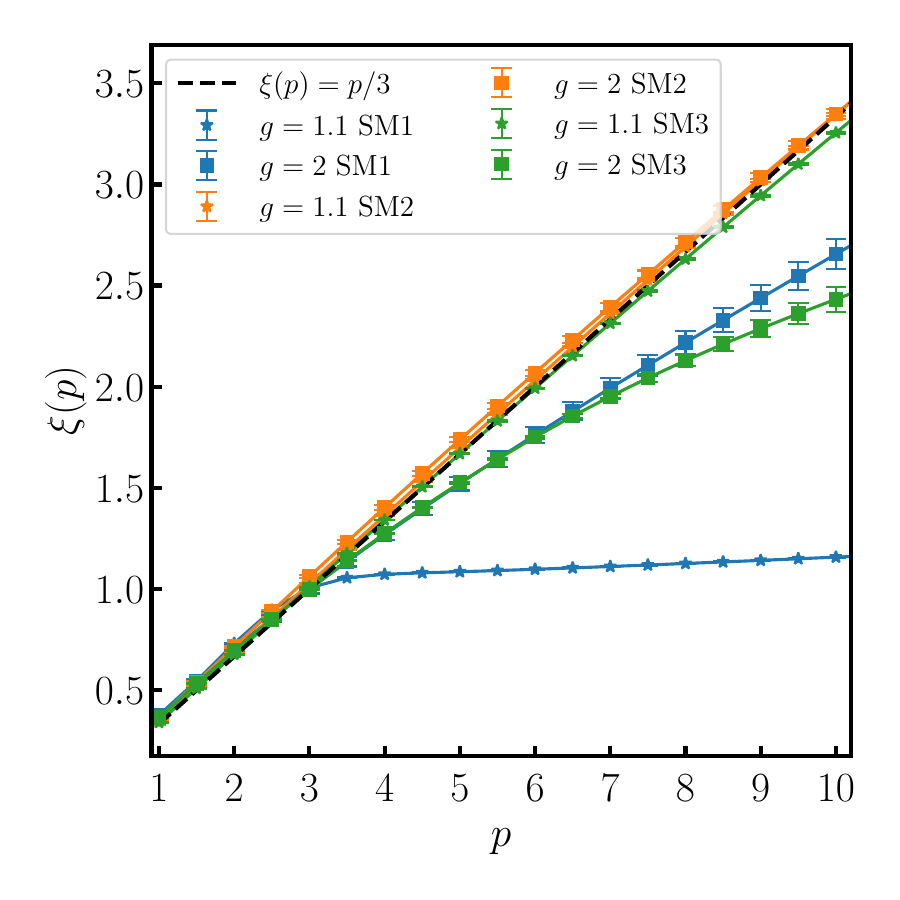}\caption{\protect\label{fig:helical_xi_p}Plot of the Scaling exponent $\xi_{p}$of
the $p-$th order structure function as a function of $p$, for different
Helical shell models corresponding to the classes SM1, SM2, and SM3
(indicated by different colors). The different markers styles represent
the two inter-shell spacings $g=2$ and $g=1.1$. The scaling exponents
were estimated using the ESS procedure described in Sec.\ref{sec:Intermittency-1}.
The dashed line represents the Kolmogorov $p/3$ scaling.}
\end{figure}

Due to their more complicated structure, helical shell model are less
explored in literature. Here, we restrict our analysis to shell models
constructed from a single class of interactions, as in Refs. \citealp{benzi:96,depietro:15},
focusing first on those classes that can sustain a forward energy
cascade for local interactions, namely SM1, SM2 and SM3. We note that:
\begin{itemize}
\item The \emph{separable} class SM1 behaves exactly as two uncoupled GOY
models. As a result, the intermittency corrections present the same
dependence with respect to the inter-shell ratio, and its synchronization
properties are equivalent to those of a single chain, as characterized
in detail in the first part of this work, showing a the transition
from the standard intermittency ($g=2$) to strong intermittency as
the inter-shell spacing $g$ is decreased.
\item On the other hand, SM2 is known to be non intermittent for the standard
$g=2$ shell spacing, and this non-intermittent behavior is robust
under variations of $g.$
\item In stark contrast to all the examples that we have seen earlier, SM3
presents a level of intermittency comparable to SM1 for $g=2$ but
the intermittency corrections decrease drastically in the small $g$
limit, and the model tends toward exhibiting monofractal scaling \citep{benzi:1996}.
\end{itemize}
Those behaviors are presented together in Fig. \ref{fig:helical_xi_p}.

\begin{figure}
\includegraphics[width=0.9\columnwidth]{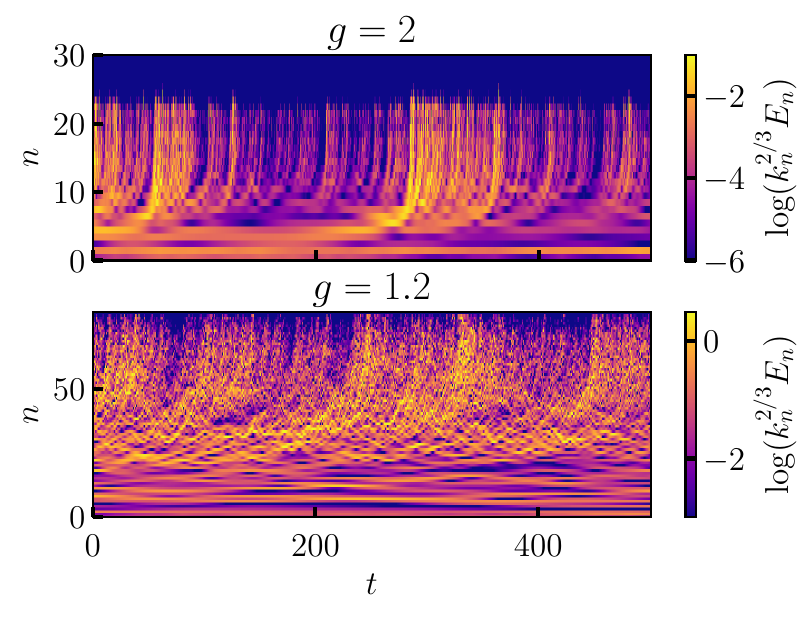}\caption{\protect\label{fig:helical_kura}Temporal evolution of compensated
energy, $log(k_{n}^{2/3}E_{n}$) across different shells (index $n$
in the $y$-axis) for the two spacing values $g=2$ and $g=1.2$ considered
(top to bottom). The plots, corresponding to the SM3 class, illustrate
the reduction of coherent structures across scales in the energy cascade
as $g$ decreases.}
\end{figure}

The time behavior of the SM3 model for the two representative cases
with inter-shell ratio $g=1.2$ and $g=2$ is presented in Fig. \ref{fig:helical_kura}
showing the evolution of the compensated energy spectrum, that can
be contrasted to Fig. \ref{fig:burstcomp-1}.

In intermittent case ($g=2$), we observe bursty fluctuations around
the global scaling $k_{n}^{-2/3}$, less defined but similar to those
seen for the GOY model evolution. However no clear signature of instantons
can be seen here. It is worth noting that, instantons have been observed
in helical shell models when initialized from a zero-background rather
than in conditions of a steady state energy cascade. In particular
for SM3 class, such structures display an energy spectrum less steep
than the Kolmogorov scaling, are reported in Ref. \citealp{depietro:2017}.

In contrast, in the non-intermittent case $(g=1.2)$, while some local
coherent events, that are carrying energy to smaller scales are still
present, they do not seem to propagate across many shells as in the
intermittent case since the compensated spectra look more noisy at
smaller scales. This suggests that triadic phases lose their ability
to self-organize in a way that accommodate strong flux events involving
a large portion of the inertial range.

The SM3 class, displays a transition opposite to that of the GOY model
(or SM1), showing a clear a reduction of the intermittency corrections
as the inter-shell ratio $g$ is reduced. This makes SM3 suitable
as a second example for investigating how the phase coherence change
moving from an intermittent to a non-intermittent regime, but with
an opposite dependence on $g$.

\begin{figure}
\includegraphics[width=0.9\columnwidth]{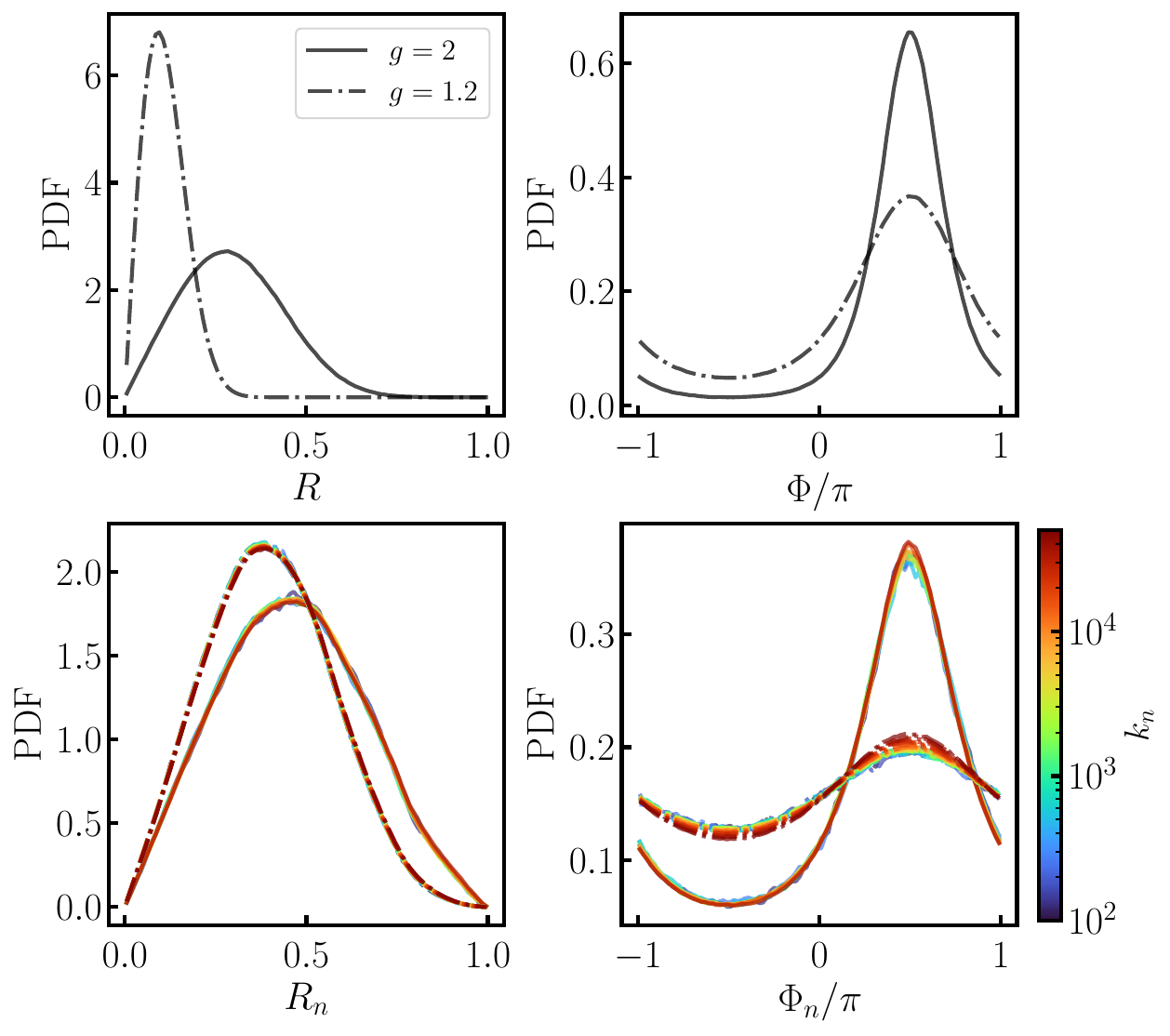}\caption{\protect\label{fig:SM3-kuravsg}The plots show the decrease in phase
coherence observed in the SM3 class as the inter-shell ratio is reduced,
for the two representative cases $g=1.2$ (dashed lines) and $g=2$
(solid lines). The top panels displays local coherence showing the
normalized PDFs of the global Kuramoto parameter $R$ (left), and
the associated average phase $\Phi$ (right). The bottom panels display
local coherence, showing the PDFs of the local Kuramoto parameter
$R_{n}$ (left) and the associated average phase $\Phi_{n}$ (right)
for different shell wave-numbers, as indicated by the colorbar.}
\end{figure}

The phase coherence in the two regimes is shown in Fig. \ref{fig:SM3-kuravsg},
where normalized PDFs of the global Kuramoto order parameter $R$,
and the associated average phase $\Phi$ are presented, for the two
inter-shell spacing $g=1.2$ and $g=2$. A clear difference in the
level of synchronization is observed, as the system moves from intermittent
to non-intermittent behavior, which is manifested by a marked reduction
in triadic phase coherence. The monofractal case exhibits a narrow
distribution at low $R$ values, in contrast to the broader distribution
of $g=2$ case, indicating that incoherent events dominate the dynamics.
At the same time, the PDF of the average triadic phase $\Phi$ appears
broader as the coherence is lost, even though a clustering around
$\pi/2$ persist, which on average, continues to sustain a forward
energy flux.

In the bottom panels, we present the PDFs of the parameters quantifying
local coherence, as defined in Eq. (\ref{eq:kurloc-1}), namely $R_{n}$
and $\Phi_{n}$, for shells within the inertial range. The local level
of phase coherence appear to be universal, i.e. independent of the
scale. Moreover, the $g=2$ case displays a distribution of local
coherence significantly different from that of the GOY model (presented
in Fig. \ref{fig:joint_pdfs-g2}), also due to the differences in
the definition of the local Kuramoto parameters in the two cases,
even though both models show comparable levels of intermittency.

Notably, local coherence persists despite the loss of global synchronization
as we can see by comparing the PDFs of $R_{n}$ in the two cases $g=2$
and $g=1.2$. When the inter-shell spacing is decreased the position
of the maxima shifts slightly toward lower coherence levels, and only
highly local coherent events ($R_{n}\simeq0,8$) are significantly
less probable. In contrast, the statistics of both local and global
mean triadic phase $\Phi$ show that the phase distribution is more
coherent for larger inter-shell spacing. This statistical features
are consistent with the observed temporal dynamics, characterized
not by structures that travels over many shells but by events that
involve neighboring scales.

Regarding forward cascades, we conclude from these results that the
Kuramoto parameter, defined using the triadic phases, provides a useful
order parameter for studying phase coherence and intermittency in
a network of interacting triads. However, it is important to stress
that it is not a unique indicator of intermittency. For instance,
for the same inter-shell ratio $g=2$ the PDF of the Kuramoto parameter
differs between GOY model and the helical model SM3, even though they
present very similar levels of intermittency. This observation suggest
that the Kuramoto parameter is also sensible to the topological structure
of the network of interacting triads. Nonetheless, once the topology
is fixed, it proves to be a highly sensible proxy of the modification
of anomalous scaling behavior, both when increasing and decreasing
the level of intermittency.

\section{Inverse cascades\protect\label{sec:Inverse-cascades}}

Despite the success of shell models in modeling forward cascades,
attempts to extend them, in order to reproduce the inverse energy
cascade observed in 2D turbulence - simply by changing the dimensionality
of the conserved invariants - have generally failed, mainly because
of the tendency of the inverse cascade in these models to be overwhelmed
by statistical shell-equipartition solutions (see, e.g., Refs. \citealp{aurell:94b,ditlevsen:1996}).
To overcome this issue, different generalization has been proposed,
such as shell model on hierarchical trees \citep{aurell:94,aurell:97},
or models based on spiral chains written for energy \citep{gurcan:19}
or a model only for amplitudes, or one based on a stochastic cascade
mechanism \citep{ditlevsen:12}. These results suggest that the self-consistently
evolving phase equation plays a crucial role in suppressing the inverse
cascade.

In this work, we will focus on the two shell model that, to our knowledge,
are able to present a genuine inverse energy cascade: the elongated
helical shell model proposed by \citet{depietro:15} and the model
studied by \citet{tom:2017} (see also Ref. \citealp{vladimirova:2021}
for a shell model that shows an inverse cascade driven by resonantly
interacting waves in the context of wave turbulence). Our goal here
is to analyze the statistical properties of triadic phases and their
implication in the inverse cascade solutions produced by these models
as a complementary example to the forward cascade case.

Since inverse cascades are boringly non-intermittent and lack coherent
bursts, we do not show detailed investigations of the \textit{synchronizability}
of these models. However it is interesting to show the forms of the
PDFs of the triadic phases together with the fluxes and the resulting
spectra, in order to demonstrate how the triadic phases organize statistically,
to drive an inverse cascade even when they are mostly random. We hope
that, understanding how triadic phase organization works in such reduced
representations as shell models, will provide the key to explaining
how the cascade mechanism is sustained in turbulence in general, including
the inverse cascade.

\subsubsection{Inverse energy cascade in an elongated helical shell model\protect\label{subsec:Inverse-helical}}

A particular class of Helical interactions, represented by triads
of the type $(-,+,+)$ or $(+,-,-)$, i.e. the SM2, is known to drive
an inverse cascade for sufficiently elongated triads. The critical
elongation, at which the direction of energy transfer changes, (i.e.
$p/k=0.278$, assuming $p<k<q$) was first predicted by Waleffe \citep{waleffe:92},
later confirmed with an elongated helical shell model \citep{depietro:15},
and can apparently, also be justified using arguments based on the
conservation of pseudo-invariants \citep{rathmann:2017} in a single
triad.

In this section we consider this model, representing a unique example
of an inverse cascade driven by non local interactions as our test
bed for studying the phase coherence of elongated triads, which drive
the inverse cascade process. We recall that for non-local interactions
the relevant triadic phases centered at the $n$-th shell can be defined
as $u_{n-l}^{s_{1}}u_{n}^{s_{0}}u_{n+m}^{s_{2}}=\chi_{n}e^{i\varphi_{n}}$
and $u_{n-l}^{-s_{1}}u_{n}^{-s_{0}}u_{n+m}^{-s_{2}}=\bar{\chi}_{n}e^{i\bar{\varphi}_{n}}$.

\begin{figure}
\includegraphics[width=0.8\columnwidth]{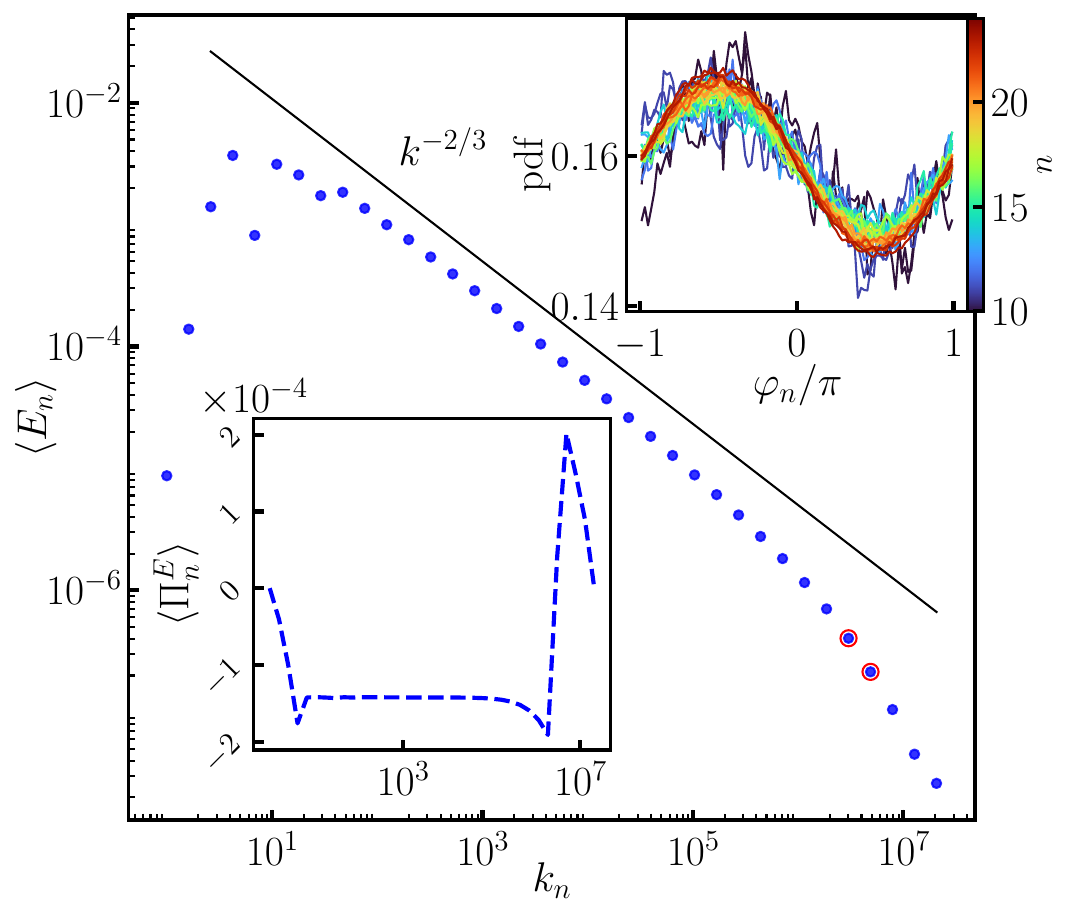}\caption{\protect\label{fig:inv-helical}Plot of the energy spectrum as a function
of the wave number for a elongated helical shell model (SM2), driven
by small-scale forcing (forcing shells indicated in red). The inset
plot on the bottom left shows the spectral flux as a function of scale,
corresponding to an inverse energy flux in the inertial range. The
inset plot on the top right shows the normalized PDFs of triadic phases
$\varphi_{n}$, for the shells in the inertial range. The shell number
is represented by different colors labeled by the continuous colorbar.}
\end{figure}

In Figure (\ref{fig:inv-helical}) we present the numerical results
for SM2 (elongated) shell model, defined by $s_{1}/s_{0}=-1$ and
$s_{2}/s_{0}=+1$, with an inter-shell ratio $g=\frac{1+\sqrt{5}}{2}$
and sufficiently elongated triads, specifically $l=3$ and $m=1$,
which satisfy $g^{-l}<0.278$. With this choice of parameters, the
introduction of a small scale forcing produces an inverse cascade,
as evidenced by the spectral energy flux and the presence of a clear
inertial range, with an energy spectrum consistent with the Kolmogorov
scaling. The probability distribution functions of non-local triadic
phases, presented in the inset (top right), show a higher probability
of events with triadic phases equal to $-\pi/2$, which corresponds
to the choice that maximizes an inverse energy flux. The distribution
appears to have a roughly sinusoidal shape and seems to be universal,
in the sense that it does not seem to show any scaling with the shell
number $n$.

Comparing these PDFs with the forward cascade case, one finds a lower
probability of coherent events in the inverse cascade. This weaker
phase coherence is even more pronounced at a global level as shown
in Figure \ref{fig:kura-cascades}, where the Kuramoto order parameter
reveals a relatively incoherent cascade dynamics, characterized by
low values of $R$, and a clear flattening of the average phase $\Phi$
distribution.

\subsubsection{Inverse energy cascades and equipartition solutions in a local shell
model\protect\label{subsec:inverse-ray-model}}

For the Sabra shell model \citep{lvov:98}, a detailed phase diagram
was proposed in Ref. \citealp{gilbert:2002}, and a further investigations
of the equipartition and the inverse cascade solutions have been carried
out by \citet{tom:2017}. Specifically, they considered $g=2$, by
setting $a=1$ and imposing energy conservation through the condition
$c=-1-b$. This causes the second positive definite invariant (in
the range $-2<b<-1$) to have the form $H=\sum_{n}k_{n}^{\alpha}|u_{n}|^{2}\;$
, with a scaling exponent $\alpha=log_{g}\left(\left|\frac{1}{-1-b}\right|\right)$.
Notably, the choice of enstrophy conservation, i.e. $\alpha=2$, corresponds
to $b=-5/4$ but as we already know this choice fails to drive an
inverse cascade because of the issue with the equipartition in shell
models. However, they showed that this shell model sustains an inverse
energy cascade for values of $b$ in the range $-2<b<b_{c}=-1.63$,
with a spectral slope that approaches the characteristic scaling of
2D turbulence in the limit $b\to-2$. We have replicated the numerical
results presented in Ref. \citealp{tom:2017}, so that we can investigate
the role of phase coherence in driving a local inverse cascade.

\begin{figure}
\includegraphics[width=0.8\columnwidth]{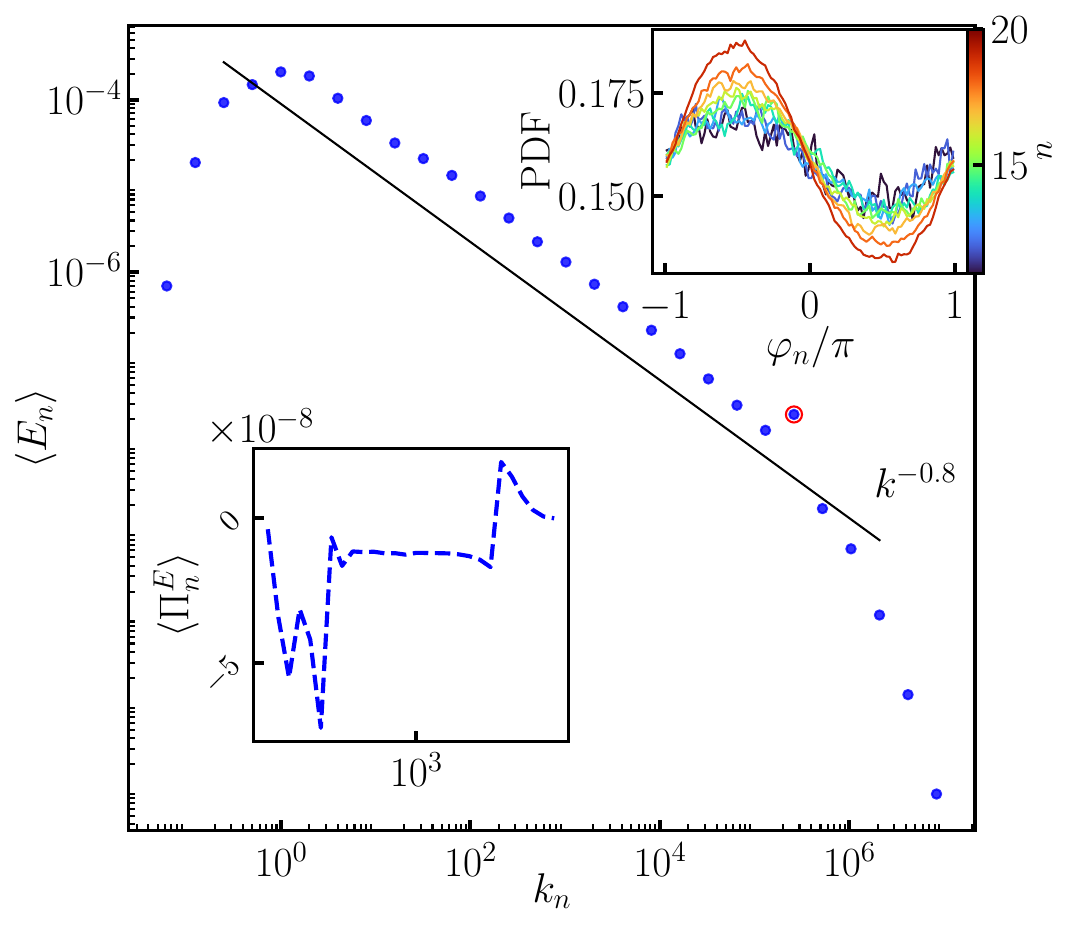}\caption{\protect\label{fig:inverse_ray}Plot of the energy spectrum as a function
of the wave number $k_{n}$, for the model of Tom and Ray, driven
by a small-scale forcing (forcing shell indicated in red). The bottom
left inset plot shows the spectral flux as a function of scale, corresponding
to an inverse energy cascade. The plotted spectral slope $k_{n}^{-0.8}$
corresponds to the numerical estimate from previous studies cited
above. The top-right inset shows the normalized PDFs of triad phases
$\varphi_{n}$ for the shells in the inertial range. The shell number
is represented by different colors labeled by the continuous colorbar.}
\end{figure}

As shown in Figure \ref{fig:inverse_ray}, the model, when forced
at small scales with $b=-1.8$, exhibits a clear inertial range associated
with an inverse energy cascade, corresponding to a constant, on average,
negative energy flux. The probability distribution of triadic phases,
displayed in the top right inset, shows how the triadic phases for
different shells organize around the value $-\pi/2,$ which maximizes
the inverse energy transfer. The distribution follows a sinusoidal
shape, i.e odd with respect to $\varphi_{n}\to-\varphi_{n}$, with
non universal, scale-dependent behavior, indicating that triadic phase
organization is weaker, (as marked by noisier and broader curves)
going to larger scales (slower time scales). In contrast (not shown),
when the model exhibits energy equipartition (i.e. for \textbf{$b_{c}<b<-1$})
triadic phases appear noisy and randomly distributed.

To quantify synchronization at a global level, providing an overall
view of phase coherence between forward and inverse cascades, we report
in Figure \ref{fig:kura-cascades}, the global coherence order parameter
$R$ and the associated average triadic phase $\Phi$. Notably, we
observe that inverse cascades are generally driven by much weaker
phase organization. Among the two inverse cascade cases considered,
the local inverse cascade displays higher level of phase coherence
compared to the one of the helical shell model driven by elongated
triads. In the latter case, even though the average phase $\Phi$
appears flattened, a backward flux persists.

\section{\protect\label{sec:Conclusion}Conclusion}

A detailed analysis of the dynamics of triadic phases in several shell
models, interpreted as chains of interacting triads, where each triad
is considered as a \emph{nonlinear oscillator}, reveals how \textit{synchronizability}
plays a key role in the statistical properties of the energy cascade.
In particular, our results show that phase-synchronized events are
associated with bursts of energy transfer, which determine the intermittency
of the system. Our investigation, which is in line with previous results
on Burgers equation that links intermittency to phase dynamics, provides
new evidence that stronger intermittency is related to enhanced triadic
phase coherence, which is demonstrated first in a shell model representation
of the three dimensional Navier Stokes equations.

In the classical GOY model, with fixed inter-shell spacing, $g=2$,
the system exhibits the ``usual level'' of intermittency. Decreasing
inter-shell spacing below a critical value, while keeping the same
physical quantities conserved, we systematically examine how each
scaling exponent changes continuously, ultimately approaching a Burgers-like
dynamical scaling in the continuum limit. This transition is accompanied
by a loss of scale invariance in the statistics of triadic phases
and in the Kolmogorov multipliers. Remarkably, the enhancement of
intermittency correction is associated with a breaking of the statistical
symmetry of both quantities: while their PDFs are scale-independent
at the ``usual level'' of intermittency, they lose this invariance
as the intermittency increases, providing an interesting challenge
to hidden-symmetry theories \citep{mailybaev:2022,mailybaev_rev:2022}.

This progressive enhancement of intermittency, controlled by the inter-shell
spacing, allows us to investigate the connection between synchronization
and intermittency, systematically using a Kuramoto order parameter
to quantify global phase coherence among the triads of the shell model.
The PDFs of the order parameter $R$ reveal an increased likelihood
for fully coherent events as $g$ decreases, complementary to this,
the average phase $\Phi$ (i.e. the phase associated with the complex
Kuramoto parameter) distribution becomes more peaked around $\pi/2$,
which is incidentally the value that maximizes forward energy transfer.

In shell models, extreme energy flux events are associated with coherent
pulses, propagating across the inertial range. We observe that, as
$g$ decreases, those pulses becomes more localized and significantly
more intense but less frequent. In order to study the correlation
between the dynamics of such burst-like events and the phase coherence
of the triad chain, a novel order parameter was introduced, to quantify
local phase coherence. This weighted Kuramoto parameter, derived from
the right-hand side of the triadic phase evolution equation, captures
how a coherent pulse acts as a synchronization domino effect on consecutive
triadic phases. Specifically, the passage of a coherent event is preceded
by a rapid increase in the local coherence parameter that synchronizes
a cluster of triads in a partially phase locked state, i.e. $R_{n}\simeq1$,
or $\Phi_{n}\simeq\pi/2,$ that maximizes the energy transfer between
aligned triads, suggesting that the local synchronization is accompanied
by the coherent burst in energy transfer. The strong correlation between
this phase locked state and large deviations in energy flux, can also
be evidenced by their joint statistics.

The PDFs of the local Kuramoto parameter also show that local coherence
precedes global synchronization. As intermittency increases, the coherent
events generate longer-lasting phase-locked states, as demonstrated
by the elevated probability of observing $R_{n}\simeq1$. This suggests
that accumulation of energy at large scales due to an incoherent phase
distribution at small scales, triggers the instanton, whose forced
passage reorganizes the phases in such a way that the energy can continue
to flow along this opened ``path'' for a while, which is then slowly
desynchronized from small scales, and as the desynchronization propagates
to large scales and becomes significant, the ``path'' gets closed
again, which in turn causes the accumulation of energy at large scales,
eventually repeating the whole cycle.

As these observations show, the local Kuramoto parameter can be a
useful proxy for detecting coherent events that drive the energy cascade
across range of scales. It may be possible to generalize this for
detecting paths of burst events to models with spatial structure such
as binary trees or more complex $k$-space structure such as nested
polyhedra models, and maybe eventually to direct numerical simulations.
In particular, the analysis of tradic phase dynamics in a 3D DNS could
follow two different approaches: focusing on a single ``path'' in
Fourier space, represented by a self-similar chain of connected triads
in the spirit of spiral-chain models, or filtering DNS data into a
set of shells and studying the triadic phase dynamics of those. However,
it should be mentioned that defining a local phase coherence indicator
in nonlinear systems involving many interconnected triads is a nontrivial
task.

As a minimal example of the case with interconnected triads, we investigated
triadic phase dynamics in helical shell models constructed from two
sequences of helical variables, allowing for different classes of
chiral interactions, focusing in particular on the SM3 class, which
consist of two intertwined chains that are topologically \textit{non-separable}.
We observe that, contrary to the single chain case, here, decreasing
the inter-shell spacing leads to a reduction of intermittency, with
the system exhibiting self-similar monofractal scaling as it approaches
the continuum limit. However, as before we find that the phase coherence
correlates with intermittency, as shown by the order parameters, even
though the dependence of that with $g$ is the inverse of what happens
in the GOY case.

Finally, we have addressed the role of triadic phases in the inverse
energy cascade, considering two specific cases, represented by a non-local
helical shell model and the model investigated in Ref. \citealp{tom:2017}.
Our results show that inverse cascades are characterized by a PDF
of triadic phases, peaked around $\varphi_{n}=-\pi/2$ , which maximizes
the inverse energy transfer. However the order parameters show that
the inverse cascades (at least the ones that we considered), are non-intermittent,
and present significantly lower levels of phase coherence in contrast
to the forward cascades. This is also in agreement with the emerging
picture that phase dynamics and intermittency in cascade models are
intimately connected.

In short, our findings provide new evidence that phase synchronization
acts as a driver of intermittency and extreme events in shell models.
We introduced novel parameters to quantify the local phase coherence
in these models, which can be extended to similar cascade models using
a network perspective, and eventually to direct numerical simulations.

\section*{ACKNOWLEDGMENT}

This work has benefited from a grant managed by the Agence Nationale
de la Recherche (ANR), as part of the program \textquoteleft Investissements
d\textquoteright Avenir\textquoteright{} under the reference (ANR-18-EURE-0014)
and has been carried out within the framework of the EUROfusion Consortium,
funded by the European Union via the Euratom Research and Training
Programme (Grant Agreement No 101052200 --- EUROfusion) and within
the framework of the French Research Federation for Fusion Studies.
The authors would like to thank the Isaac Newton Institute for Mathematical
Sciences, Cambridge, for support and hospitality during the programme
``Anti-diffusive dynamics: from sub-cellular to astrophysical scales'',
where part of the work on this paper was undertaken. This work was
supported by EPSRC grant EP/R014604/1. In particular, the authors
would like to thank Dr. Santiago J. Benavides for stimulating discussion
on phase dynamics during the program.

\section*{DATA AVAILABILITY}

The data that support the findings of this article are openly available
at \citep{manfredinidataset:2025}.

\bibliographystyle{aipnum4-2}

\begin{thebibliography}{74}%
\makeatletter
\providecommand \@ifxundefined [1]{%
 \@ifx{#1\undefined}
}%
\providecommand \@ifnum [1]{%
 \ifnum #1\expandafter \@firstoftwo
 \else \expandafter \@secondoftwo
 \fi
}%
\providecommand \@ifx [1]{%
 \ifx #1\expandafter \@firstoftwo
 \else \expandafter \@secondoftwo
 \fi
}%
\providecommand \natexlab [1]{#1}%
\providecommand \enquote  [1]{``#1''}%
\providecommand \bibnamefont  [1]{#1}%
\providecommand \bibfnamefont [1]{#1}%
\providecommand \citenamefont [1]{#1}%
\providecommand \href@noop [0]{\@secondoftwo}%
\providecommand \href [0]{\begingroup \@sanitize@url \@href}%
\providecommand \@href[1]{\@@startlink{#1}\@@href}%
\providecommand \@@href[1]{\endgroup#1\@@endlink}%
\providecommand \@sanitize@url [0]{\catcode `\\12\catcode `\$12\catcode
  `\&12\catcode `\#12\catcode `\^12\catcode `\_12\catcode `\%12\relax}%
\providecommand \@@startlink[1]{}%
\providecommand \@@endlink[0]{}%
\providecommand \url  [0]{\begingroup\@sanitize@url \@url }%
\providecommand \@url [1]{\endgroup\@href {#1}{\urlprefix }}%
\providecommand \urlprefix  [0]{URL }%
\providecommand \Eprint [0]{\href }%
\providecommand \doibase [0]{https://doi.org/}%
\providecommand \selectlanguage [0]{\@gobble}%
\providecommand \bibinfo  [0]{\@secondoftwo}%
\providecommand \bibfield  [0]{\@secondoftwo}%
\providecommand \translation [1]{[#1]}%
\providecommand \BibitemOpen [0]{}%
\providecommand \bibitemStop [0]{}%
\providecommand \bibitemNoStop [0]{.\EOS\space}%
\providecommand \EOS [0]{\spacefactor3000\relax}%
\providecommand \BibitemShut  [1]{\csname bibitem#1\endcsname}%
\let\auto@bib@innerbib\@empty
%</preamble>
\bibitem [{\citenamefont {Bak}\ \emph {et~al.}(2002)\citenamefont {Bak},
  \citenamefont {Christensen}, \citenamefont {Danon},\ and\ \citenamefont
  {Scanlon}}]{bak:2002}%
  \BibitemOpen
  \bibfield  {author} {\bibinfo {author} {\bibfnamefont {P.}~\bibnamefont
  {Bak}}, \bibinfo {author} {\bibfnamefont {K.}~\bibnamefont {Christensen}},
  \bibinfo {author} {\bibfnamefont {L.}~\bibnamefont {Danon}},\ and\ \bibinfo
  {author} {\bibfnamefont {T.}~\bibnamefont {Scanlon}},\ }\href
  {https://doi.org/10.1103/PhysRevLett.88.178501} {\bibfield  {journal}
  {\bibinfo  {journal} {Phys. Rev. Lett.}\ }\textbf {\bibinfo {volume} {88}},\
  \bibinfo {pages} {178501} (\bibinfo {year} {2002})}\BibitemShut {NoStop}%
\bibitem [{\citenamefont {Gopikrishnan}\ \emph {et~al.}(1999)\citenamefont
  {Gopikrishnan}, \citenamefont {Plerou}, \citenamefont {Nunes~Amaral},
  \citenamefont {Meyer},\ and\ \citenamefont {Stanley}}]{gopikrishnan:1999}%
  \BibitemOpen
  \bibfield  {author} {\bibinfo {author} {\bibfnamefont {P.}~\bibnamefont
  {Gopikrishnan}}, \bibinfo {author} {\bibfnamefont {V.}~\bibnamefont
  {Plerou}}, \bibinfo {author} {\bibfnamefont {L.~A.}\ \bibnamefont
  {Nunes~Amaral}}, \bibinfo {author} {\bibfnamefont {M.}~\bibnamefont
  {Meyer}},\ and\ \bibinfo {author} {\bibfnamefont {H.~E.}\ \bibnamefont
  {Stanley}},\ }\href {https://doi.org/10.1103/PhysRevE.60.5305} {\bibfield
  {journal} {\bibinfo  {journal} {Phys. Rev. E}\ }\textbf {\bibinfo {volume}
  {60}},\ \bibinfo {pages} {5305} (\bibinfo {year} {1999})}\BibitemShut
  {NoStop}%
\bibitem [{\citenamefont {Lovejoy}\ and\ \citenamefont
  {Schertzer}(2013)}]{lovejoy:2013}%
  \BibitemOpen
  \bibfield  {author} {\bibinfo {author} {\bibfnamefont {S.}~\bibnamefont
  {Lovejoy}}\ and\ \bibinfo {author} {\bibfnamefont {D.}~\bibnamefont
  {Schertzer}},\ }\href@noop {} {\emph {\bibinfo {title} {The weather and
  climate: emergent laws and multifractal cascades}}}\ (\bibinfo  {publisher}
  {Cambridge University Press},\ \bibinfo {year} {2013})\BibitemShut {NoStop}%
\bibitem [{\citenamefont {Frisch}(1995)}]{frisch}%
  \BibitemOpen
  \bibfield  {author} {\bibinfo {author} {\bibfnamefont {U.}~\bibnamefont
  {Frisch}},\ }\href@noop {} {\emph {\bibinfo {title} {Turbulence: The Legacy
  of A. N. Kolmogorov}}}\ (\bibinfo  {publisher} {Cambridge University Press},\
  \bibinfo {address} {Cambridge},\ \bibinfo {year} {1995})\BibitemShut
  {NoStop}%
\bibitem [{\citenamefont {Bohr}\ \emph {et~al.}(1998)\citenamefont {Bohr},
  \citenamefont {Jensen}, \citenamefont {Paladin},\ and\ \citenamefont
  {Vulpiani}}]{bohr_Jensen_Paladin_Vulpiani_1998}%
  \BibitemOpen
  \bibfield  {author} {\bibinfo {author} {\bibfnamefont {T.}~\bibnamefont
  {Bohr}}, \bibinfo {author} {\bibfnamefont {M.~H.}\ \bibnamefont {Jensen}},
  \bibinfo {author} {\bibfnamefont {G.}~\bibnamefont {Paladin}},\ and\ \bibinfo
  {author} {\bibfnamefont {A.}~\bibnamefont {Vulpiani}},\ }\href@noop {} {\emph
  {\bibinfo {title} {Dynamical Systems Approach to Turbulence}}},\ Cambridge
  Nonlinear Science Series\ (\bibinfo  {publisher} {Cambridge University
  Press},\ \bibinfo {year} {1998})\BibitemShut {NoStop}%
\bibitem [{\citenamefont {Biferale}(2003)}]{biferale:03}%
  \BibitemOpen
  \bibfield  {author} {\bibinfo {author} {\bibfnamefont {L.}~\bibnamefont
  {Biferale}},\ }\href {https://doi.org/10.1146/annurev.fluid.35.101101.161122}
  {\bibfield  {journal} {\bibinfo  {journal} {Ann. Rev. Fluid Mech.}\ }\textbf
  {\bibinfo {volume} {35}},\ \bibinfo {pages} {441} (\bibinfo {year}
  {2003})}\BibitemShut {NoStop}%
\bibitem [{\citenamefont {Ditlevsen}(2010)}]{ditlevsen:2010}%
  \BibitemOpen
  \bibfield  {author} {\bibinfo {author} {\bibfnamefont {P.~D.}\ \bibnamefont
  {Ditlevsen}},\ }\href@noop {} {\emph {\bibinfo {title} {Turbulence and Shell
  Models}}}\ (\bibinfo  {publisher} {Cambridge University Press},\ \bibinfo
  {year} {2010})\BibitemShut {NoStop}%
\bibitem [{\citenamefont {Hattori}, \citenamefont {Rubinstein},\ and\
  \citenamefont {Ishizawa}(2004)}]{hattori:04}%
  \BibitemOpen
  \bibfield  {author} {\bibinfo {author} {\bibfnamefont {Y.}~\bibnamefont
  {Hattori}}, \bibinfo {author} {\bibfnamefont {R.}~\bibnamefont
  {Rubinstein}},\ and\ \bibinfo {author} {\bibfnamefont {A.}~\bibnamefont
  {Ishizawa}},\ }\href@noop {} {\bibfield  {journal} {\bibinfo  {journal}
  {Phys. Rev. E}\ }\textbf {\bibinfo {volume} {70}},\ \bibinfo {pages} {046311}
  (\bibinfo {year} {2004})}\BibitemShut {NoStop}%
\bibitem [{\citenamefont {Jensen}, \citenamefont {Paladin},\ and\ \citenamefont
  {Vulpiani}(1992)}]{jensen1992}%
  \BibitemOpen
  \bibfield  {author} {\bibinfo {author} {\bibfnamefont {M.~H.}\ \bibnamefont
  {Jensen}}, \bibinfo {author} {\bibfnamefont {G.}~\bibnamefont {Paladin}},\
  and\ \bibinfo {author} {\bibfnamefont {A.}~\bibnamefont {Vulpiani}},\ }\href
  {https://doi.org/10.1103/PhysRevA.45.7214} {\bibfield  {journal} {\bibinfo
  {journal} {Phys. Rev. A}\ }\textbf {\bibinfo {volume} {45}},\ \bibinfo
  {pages} {7214} (\bibinfo {year} {1992})}\BibitemShut {NoStop}%
\bibitem [{\citenamefont {Kumar}\ and\ \citenamefont
  {Verma}(2015)}]{kumar:2015}%
  \BibitemOpen
  \bibfield  {author} {\bibinfo {author} {\bibfnamefont {A.}~\bibnamefont
  {Kumar}}\ and\ \bibinfo {author} {\bibfnamefont {M.~K.}\ \bibnamefont
  {Verma}},\ }\href {https://doi.org/10.1103/PhysRevE.91.043014} {\bibfield
  {journal} {\bibinfo  {journal} {Phys. Rev. E}\ }\textbf {\bibinfo {volume}
  {91}},\ \bibinfo {pages} {043014} (\bibinfo {year} {2015})}\BibitemShut
  {NoStop}%
\bibitem [{\citenamefont {Plunian}, \citenamefont {Stepanov},\ and\
  \citenamefont {Frick}(2013)}]{plunian:2013}%
  \BibitemOpen
  \bibfield  {author} {\bibinfo {author} {\bibfnamefont {F.}~\bibnamefont
  {Plunian}}, \bibinfo {author} {\bibfnamefont {R.}~\bibnamefont {Stepanov}},\
  and\ \bibinfo {author} {\bibfnamefont {P.}~\bibnamefont {Frick}},\ }\href
  {https://doi.org/10.1016/j.physrep.2012.09.001} {\bibfield  {journal}
  {\bibinfo  {journal} {Phys. Rep. - Rev. Sect. Phys. Lett.}\ }\textbf
  {\bibinfo {volume} {523}},\ \bibinfo {pages} {1} (\bibinfo {year}
  {2013})}\BibitemShut {NoStop}%
\bibitem [{\citenamefont {Verdini}\ and\ \citenamefont
  {Grappin}(2012)}]{verdini:12}%
  \BibitemOpen
  \bibfield  {author} {\bibinfo {author} {\bibfnamefont {A.}~\bibnamefont
  {Verdini}}\ and\ \bibinfo {author} {\bibfnamefont {R.}~\bibnamefont
  {Grappin}},\ }\href@noop {} {\bibfield  {journal} {\bibinfo  {journal} {Phys.
  Rev. Lett.}\ }\textbf {\bibinfo {volume} {109}},\ \bibinfo {pages} {025004}
  (\bibinfo {year} {2012})}\BibitemShut {NoStop}%
\bibitem [{\citenamefont {Berionni}, \citenamefont {Morel},\ and\ \citenamefont
  {G\"urcan}(2017)}]{berionni:17}%
  \BibitemOpen
  \bibfield  {author} {\bibinfo {author} {\bibfnamefont {V.}~\bibnamefont
  {Berionni}}, \bibinfo {author} {\bibfnamefont {P.}~\bibnamefont {Morel}},\
  and\ \bibinfo {author} {\bibfnamefont {{\"O}.~D.}\ \bibnamefont {G\"urcan}},\
  }\href {https://doi.org/10.1063/1.4998569} {\bibfield  {journal} {\bibinfo
  {journal} {Physics of Plasmas}\ }\textbf {\bibinfo {volume} {24}},\ \bibinfo
  {pages} {122310} (\bibinfo {year} {2017})}\BibitemShut {NoStop}%
\bibitem [{\citenamefont {Ghantous}\ and\ \citenamefont
  {G\"urcan}(2015)}]{ghantous:15}%
  \BibitemOpen
  \bibfield  {author} {\bibinfo {author} {\bibfnamefont {K.}~\bibnamefont
  {Ghantous}}\ and\ \bibinfo {author} {\bibfnamefont {{\"O}.~D.}\ \bibnamefont
  {G\"urcan}},\ }\href {https://doi.org/10.1103/PhysRevE.92.033107} {\bibfield
  {journal} {\bibinfo  {journal} {Phys. Rev. E}\ }\textbf {\bibinfo {volume}
  {92}},\ \bibinfo {pages} {033107} (\bibinfo {year} {2015})}\BibitemShut
  {NoStop}%
\bibitem [{\citenamefont {Kadanoff}\ \emph {et~al.}(1995)\citenamefont
  {Kadanoff}, \citenamefont {Lohse}, \citenamefont {Wang},\ and\ \citenamefont
  {Benzi}}]{kadanoff1995}%
  \BibitemOpen
  \bibfield  {author} {\bibinfo {author} {\bibfnamefont {L.}~\bibnamefont
  {Kadanoff}}, \bibinfo {author} {\bibfnamefont {D.}~\bibnamefont {Lohse}},
  \bibinfo {author} {\bibfnamefont {J.}~\bibnamefont {Wang}},\ and\ \bibinfo
  {author} {\bibfnamefont {R.}~\bibnamefont {Benzi}},\ }\href
  {https://doi.org/10.1063/1.868775} {\bibfield  {journal} {\bibinfo  {journal}
  {Phys. Fluids}\ }\textbf {\bibinfo {volume} {7}},\ \bibinfo {pages} {617}
  (\bibinfo {year} {1995})}\BibitemShut {NoStop}%
\bibitem [{\citenamefont {Ohkitani}\ and\ \citenamefont
  {Yamada}(1989)}]{ohkitani:89}%
  \BibitemOpen
  \bibfield  {author} {\bibinfo {author} {\bibfnamefont {K.}~\bibnamefont
  {Ohkitani}}\ and\ \bibinfo {author} {\bibfnamefont {M.}~\bibnamefont
  {Yamada}},\ }\href {https://doi.org/10.1143/PTP.81.329} {\bibfield  {journal}
  {\bibinfo  {journal} {Progress of Theoretical Physics}\ }\textbf {\bibinfo
  {volume} {81}},\ \bibinfo {pages} {329} (\bibinfo {year} {1989})}\BibitemShut
  {NoStop}%
\bibitem [{\citenamefont {L\char39{}vov}\ \emph {et~al.}(1998)\citenamefont
  {L\char39{}vov}, \citenamefont {Podivilov}, \citenamefont {Pomyalov},
  \citenamefont {Procaccia},\ and\ \citenamefont {Vandembroucq}}]{lvov:98}%
  \BibitemOpen
  \bibfield  {author} {\bibinfo {author} {\bibfnamefont {V.~S.}\ \bibnamefont
  {L\char39{}vov}}, \bibinfo {author} {\bibfnamefont {E.}~\bibnamefont
  {Podivilov}}, \bibinfo {author} {\bibfnamefont {A.}~\bibnamefont {Pomyalov}},
  \bibinfo {author} {\bibfnamefont {I.}~\bibnamefont {Procaccia}},\ and\
  \bibinfo {author} {\bibfnamefont {D.}~\bibnamefont {Vandembroucq}},\
  }\href@noop {} {\bibfield  {journal} {\bibinfo  {journal} {Phys. Rev. E}\
  }\textbf {\bibinfo {volume} {58}},\ \bibinfo {pages} {1811} (\bibinfo {year}
  {1998})}\BibitemShut {NoStop}%
\bibitem [{\citenamefont {L'vov}, \citenamefont {Pomyalov},\ and\ \citenamefont
  {Procaccia}(2001)}]{lvov:2001}%
  \BibitemOpen
  \bibfield  {author} {\bibinfo {author} {\bibfnamefont {V.~S.}\ \bibnamefont
  {L'vov}}, \bibinfo {author} {\bibfnamefont {A.}~\bibnamefont {Pomyalov}},\
  and\ \bibinfo {author} {\bibfnamefont {I.}~\bibnamefont {Procaccia}},\ }\href
  {https://doi.org/10.1103/PhysRevE.63.056118} {\bibfield  {journal} {\bibinfo
  {journal} {Phys. Rev. E}\ }\textbf {\bibinfo {volume} {63}},\ \bibinfo
  {pages} {056118} (\bibinfo {year} {2001})}\BibitemShut {NoStop}%
\bibitem [{\citenamefont {de~Wit}\ \emph {et~al.}(2024)\citenamefont {de~Wit},
  \citenamefont {Ortali}, \citenamefont {Corbetta}, \citenamefont {Mailybaev},
  \citenamefont {Biferale},\ and\ \citenamefont {Toschi}}]{de-wit:2024:}%
  \BibitemOpen
  \bibfield  {author} {\bibinfo {author} {\bibfnamefont {X.~M.}\ \bibnamefont
  {de~Wit}}, \bibinfo {author} {\bibfnamefont {G.}~\bibnamefont {Ortali}},
  \bibinfo {author} {\bibfnamefont {A.}~\bibnamefont {Corbetta}}, \bibinfo
  {author} {\bibfnamefont {A.~A.}\ \bibnamefont {Mailybaev}}, \bibinfo {author}
  {\bibfnamefont {L.}~\bibnamefont {Biferale}},\ and\ \bibinfo {author}
  {\bibfnamefont {F.}~\bibnamefont {Toschi}},\ }\href
  {https://doi.org/10.1103/PhysRevE.109.055106} {\bibfield  {journal} {\bibinfo
   {journal} {Phys. Rev. E}\ }\textbf {\bibinfo {volume} {109}},\ \bibinfo
  {pages} {055106} (\bibinfo {year} {2024})}\BibitemShut {NoStop}%
\bibitem [{\citenamefont {Mailybaev}(2012)}]{mailybaev:12}%
  \BibitemOpen
  \bibfield  {author} {\bibinfo {author} {\bibfnamefont {A.~A.}\ \bibnamefont
  {Mailybaev}},\ }\href {https://doi.org/10.1103/PhysRevE.86.025301} {\bibfield
   {journal} {\bibinfo  {journal} {Phys. Rev. E}\ }\textbf {\bibinfo {volume}
  {86}},\ \bibinfo {pages} {025301(R)} (\bibinfo {year} {2012})}\BibitemShut
  {NoStop}%
\bibitem [{\citenamefont {Mailybaev}(2013)}]{mailybaev:2013}%
  \BibitemOpen
  \bibfield  {author} {\bibinfo {author} {\bibfnamefont {A.~A.}\ \bibnamefont
  {Mailybaev}},\ }\href {https://doi.org/10.1103/PhysRevE.87.053011} {\bibfield
   {journal} {\bibinfo  {journal} {Phys. Rev. E}\ }\textbf {\bibinfo {volume}
  {87}},\ \bibinfo {pages} {053011} (\bibinfo {year} {2013})}\BibitemShut
  {NoStop}%
\bibitem [{\citenamefont {G\"urcan}(2017)}]{gurcan:17}%
  \BibitemOpen
  \bibfield  {author} {\bibinfo {author} {\bibfnamefont {{\"O}.~D.}\
  \bibnamefont {G\"urcan}},\ }\href
  {https://doi.org/10.1103/PhysRevE.95.063102} {\bibfield  {journal} {\bibinfo
  {journal} {Phys. Rev. E}\ }\textbf {\bibinfo {volume} {95}},\ \bibinfo
  {pages} {063102} (\bibinfo {year} {2017})}\BibitemShut {NoStop}%
\bibitem [{\citenamefont {G\"urcan}\ \emph {et~al.}(2016)\citenamefont
  {G\"urcan}, \citenamefont {Morel}, \citenamefont {Kobayashi}, \citenamefont
  {Singh}, \citenamefont {Xu},\ and\ \citenamefont {Diamond}}]{gurcan:16a}%
  \BibitemOpen
  \bibfield  {author} {\bibinfo {author} {\bibfnamefont {{\"O}.~D.}\
  \bibnamefont {G\"urcan}}, \bibinfo {author} {\bibfnamefont {P.}~\bibnamefont
  {Morel}}, \bibinfo {author} {\bibfnamefont {S.}~\bibnamefont {Kobayashi}},
  \bibinfo {author} {\bibfnamefont {R.}~\bibnamefont {Singh}}, \bibinfo
  {author} {\bibfnamefont {S.}~\bibnamefont {Xu}},\ and\ \bibinfo {author}
  {\bibfnamefont {P.~H.}\ \bibnamefont {Diamond}},\ }\href
  {https://doi.org/10.1103/PhysRevE.94.033106} {\bibfield  {journal} {\bibinfo
  {journal} {Phys. Rev. E}\ }\textbf {\bibinfo {volume} {94}},\ \bibinfo
  {pages} {033106} (\bibinfo {year} {2016})}\BibitemShut {NoStop}%
\bibitem [{\citenamefont {G\"urcan}, \citenamefont {Xu},\ and\ \citenamefont
  {Morel}(2019)}]{gurcan:19}%
  \BibitemOpen
  \bibfield  {author} {\bibinfo {author} {\bibfnamefont {{\"O}.~D.}\
  \bibnamefont {G\"urcan}}, \bibinfo {author} {\bibfnamefont {S.}~\bibnamefont
  {Xu}},\ and\ \bibinfo {author} {\bibfnamefont {P.}~\bibnamefont {Morel}},\
  }\href {https://doi.org/10.1103/PhysRevE.100.043113} {\bibfield  {journal}
  {\bibinfo  {journal} {Phys. Rev. E}\ }\textbf {\bibinfo {volume} {100}},\
  \bibinfo {pages} {043113} (\bibinfo {year} {2019})}\BibitemShut {NoStop}%
\bibitem [{\citenamefont {Campolina}\ and\ \citenamefont
  {Mailybaev}(2018)}]{campolina:2018}%
  \BibitemOpen
  \bibfield  {author} {\bibinfo {author} {\bibfnamefont {C.~S.}\ \bibnamefont
  {Campolina}}\ and\ \bibinfo {author} {\bibfnamefont {A.~A.}\ \bibnamefont
  {Mailybaev}},\ }\href {https://doi.org/10.1103/PhysRevLett.121.064501}
  {\bibfield  {journal} {\bibinfo  {journal} {Phys. Rev. Lett.}\ }\textbf
  {\bibinfo {volume} {121}},\ \bibinfo {pages} {064501} (\bibinfo {year}
  {2018})}\BibitemShut {NoStop}%
\bibitem [{\citenamefont {Campolina}\ and\ \citenamefont
  {Mailybaev}(2021)}]{campolina:2021}%
  \BibitemOpen
  \bibfield  {author} {\bibinfo {author} {\bibfnamefont {C.~S.}\ \bibnamefont
  {Campolina}}\ and\ \bibinfo {author} {\bibfnamefont {A.~A.}\ \bibnamefont
  {Mailybaev}},\ }\href {https://doi.org/10.1088/1361-6544/abef73} {\bibfield
  {journal} {\bibinfo  {journal} {Nonlinearity}\ }\textbf {\bibinfo {volume}
  {34}},\ \bibinfo {pages} {4684} (\bibinfo {year} {2021})}\BibitemShut
  {NoStop}%
\bibitem [{\citenamefont {Costa}, \citenamefont {Barral},\ and\ \citenamefont
  {Dubrulle}(2023)}]{costa:2023}%
  \BibitemOpen
  \bibfield  {author} {\bibinfo {author} {\bibfnamefont {G.}~\bibnamefont
  {Costa}}, \bibinfo {author} {\bibfnamefont {A.}~\bibnamefont {Barral}},\ and\
  \bibinfo {author} {\bibfnamefont {B.}~\bibnamefont {Dubrulle}},\ }\href
  {https://doi.org/10.1103/PhysRevE.107.065106} {\bibfield  {journal} {\bibinfo
   {journal} {Phys. Rev. E}\ }\textbf {\bibinfo {volume} {107}} (\bibinfo
  {year} {2023}),\ 10.1103/PhysRevE.107.065106}\BibitemShut {NoStop}%
\bibitem [{\citenamefont {Lanotte}\ \emph {et~al.}(2015)\citenamefont
  {Lanotte}, \citenamefont {Benzi}, \citenamefont {Malapaka}, \citenamefont
  {Toschi},\ and\ \citenamefont {Biferale}}]{lanotte:15}%
  \BibitemOpen
  \bibfield  {author} {\bibinfo {author} {\bibfnamefont {A.~S.}\ \bibnamefont
  {Lanotte}}, \bibinfo {author} {\bibfnamefont {R.}~\bibnamefont {Benzi}},
  \bibinfo {author} {\bibfnamefont {S.~K.}\ \bibnamefont {Malapaka}}, \bibinfo
  {author} {\bibfnamefont {F.}~\bibnamefont {Toschi}},\ and\ \bibinfo {author}
  {\bibfnamefont {L.}~\bibnamefont {Biferale}},\ }\href@noop {} {\bibfield
  {journal} {\bibinfo  {journal} {Phys. Rev. Lett.}\ }\textbf {\bibinfo
  {volume} {115}},\ \bibinfo {pages} {264502} (\bibinfo {year}
  {2015})}\BibitemShut {NoStop}%
\bibitem [{\citenamefont {Buzzicotti}\ \emph
  {et~al.}(2016{\natexlab{a}})\citenamefont {Buzzicotti}, \citenamefont
  {Biferale}, \citenamefont {Frisch},\ and\ \citenamefont
  {Ray}}]{buzzicotti:2016}%
  \BibitemOpen
  \bibfield  {author} {\bibinfo {author} {\bibfnamefont {M.}~\bibnamefont
  {Buzzicotti}}, \bibinfo {author} {\bibfnamefont {L.}~\bibnamefont
  {Biferale}}, \bibinfo {author} {\bibfnamefont {U.}~\bibnamefont {Frisch}},\
  and\ \bibinfo {author} {\bibfnamefont {S.~S.}\ \bibnamefont {Ray}},\ }\href
  {https://doi.org/10.1103/PhysRevE.93.033109} {\bibfield  {journal} {\bibinfo
  {journal} {Phys. Rev. E}\ }\textbf {\bibinfo {volume} {93}} (\bibinfo {year}
  {2016}{\natexlab{a}}),\ 10.1103/PhysRevE.93.033109}\BibitemShut {NoStop}%
\bibitem [{\citenamefont {Buzzicotti}\ \emph
  {et~al.}(2016{\natexlab{b}})\citenamefont {Buzzicotti}, \citenamefont
  {Murray}, \citenamefont {Biferale},\ and\ \citenamefont
  {Bustamante}}]{buzzicotti-phase:2016}%
  \BibitemOpen
  \bibfield  {author} {\bibinfo {author} {\bibfnamefont {M.}~\bibnamefont
  {Buzzicotti}}, \bibinfo {author} {\bibfnamefont {B.~P.}\ \bibnamefont
  {Murray}}, \bibinfo {author} {\bibfnamefont {L.}~\bibnamefont {Biferale}},\
  and\ \bibinfo {author} {\bibfnamefont {M.~D.}\ \bibnamefont {Bustamante}},\
  }\href {https://doi.org/10.1140/epje/i2016-16034-5} {\bibfield  {journal}
  {\bibinfo  {journal} {Eur. Phys. J. E}\ }\textbf {\bibinfo {volume} {39}}
  (\bibinfo {year} {2016}{\natexlab{b}}),\
  10.1140/epje/i2016-16034-5}\BibitemShut {NoStop}%
\bibitem [{\citenamefont {Murray}\ and\ \citenamefont
  {Bustamante}(2018)}]{murray:2018}%
  \BibitemOpen
  \bibfield  {author} {\bibinfo {author} {\bibfnamefont {B.~P.}\ \bibnamefont
  {Murray}}\ and\ \bibinfo {author} {\bibfnamefont {M.~D.}\ \bibnamefont
  {Bustamante}},\ }\href {https://doi.org/10.1017/jfm.2018.454} {\bibfield
  {journal} {\bibinfo  {journal} {Journal of Fluid Mechanics}\ }\textbf
  {\bibinfo {volume} {850}},\ \bibinfo {pages} {624} (\bibinfo {year}
  {2018})}\BibitemShut {NoStop}%
\bibitem [{\citenamefont {Protas}, \citenamefont {Kang},\ and\ \citenamefont
  {Bustamante}(2024)}]{protas:2024}%
  \BibitemOpen
  \bibfield  {author} {\bibinfo {author} {\bibfnamefont {B.}~\bibnamefont
  {Protas}}, \bibinfo {author} {\bibfnamefont {D.}~\bibnamefont {Kang}},\ and\
  \bibinfo {author} {\bibfnamefont {M.~D.}\ \bibnamefont {Bustamante}},\ }\href
  {https://doi.org/10.1103/PhysRevE.109.055104} {\bibfield  {journal} {\bibinfo
   {journal} {Phys. Rev. E}\ }\textbf {\bibinfo {volume} {109}} (\bibinfo
  {year} {2024}),\ 10.1103/PhysRevE.109.055104}\BibitemShut {NoStop}%
\bibitem [{\citenamefont {G\"urcan}, \citenamefont {Li},\ and\ \citenamefont
  {Morel}(2020)}]{gurcan:20}%
  \BibitemOpen
  \bibfield  {author} {\bibinfo {author} {\bibfnamefont {{\"O}.~D.}\
  \bibnamefont {G\"urcan}}, \bibinfo {author} {\bibfnamefont {Y.}~\bibnamefont
  {Li}},\ and\ \bibinfo {author} {\bibfnamefont {P.}~\bibnamefont {Morel}},\
  }\href {https://doi.org/10.3390/math8040530} {\bibfield  {journal} {\bibinfo
  {journal} {Mathematics}\ }\textbf {\bibinfo {volume} {8}},\ \bibinfo {pages}
  {530} (\bibinfo {year} {2020})}\BibitemShut {NoStop}%
\bibitem [{\citenamefont {G\"urcan}(2021)}]{gurcan:21}%
  \BibitemOpen
  \bibfield  {author} {\bibinfo {author} {\bibfnamefont {{\"O}.~D.}\
  \bibnamefont {G\"urcan}},\ }\href
  {https://doi.org/https://doi.org/10.1016/j.physd.2021.132983} {\bibfield
  {journal} {\bibinfo  {journal} {Physica D: Nonlinear Phenomena}\ }\textbf
  {\bibinfo {volume} {426}},\ \bibinfo {pages} {132983} (\bibinfo {year}
  {2021})}\BibitemShut {NoStop}%
\bibitem [{\citenamefont {G\"urcan}(2023)}]{gurcan:23}%
  \BibitemOpen
  \bibfield  {author} {\bibinfo {author} {\bibfnamefont {{\"{O}}.~D.}\
  \bibnamefont {G\"urcan}},\ }\href
  {https://doi.org/10.1007/s41614-023-00122-7} {\bibfield  {journal} {\bibinfo
  {journal} {Reviews of Modern Plasma Physics}\ }\textbf {\bibinfo {volume}
  {7}},\ \bibinfo {pages} {20} (\bibinfo {year} {2023})}\BibitemShut {NoStop}%
\bibitem [{\citenamefont {Manfredini}\ and\ \citenamefont
  {G\"urcan}(2025)}]{manfredini:2025}%
  \BibitemOpen
  \bibfield  {author} {\bibinfo {author} {\bibfnamefont {L.}~\bibnamefont
  {Manfredini}}\ and\ \bibinfo {author} {\bibfnamefont {O.~D.}\ \bibnamefont
  {G\"urcan}},\ }\href {https://doi.org/10.1103/PhysRevE.111.025103} {\bibfield
   {journal} {\bibinfo  {journal} {Phys. Rev. E}\ }\textbf {\bibinfo {volume}
  {111}},\ \bibinfo {pages} {025103} (\bibinfo {year} {2025})}\BibitemShut
  {NoStop}%
\bibitem [{\citenamefont {Alexakis}\ and\ \citenamefont
  {Biferale}(2018)}]{alexakis:18}%
  \BibitemOpen
  \bibfield  {author} {\bibinfo {author} {\bibfnamefont {A.}~\bibnamefont
  {Alexakis}}\ and\ \bibinfo {author} {\bibfnamefont {L.}~\bibnamefont
  {Biferale}},\ }\href
  {https://doi.org/https://doi.org/10.1016/j.physrep.2018.08.001} {\bibfield
  {journal} {\bibinfo  {journal} {Physics Reports}\ }\textbf {\bibinfo {volume}
  {767-769}},\ \bibinfo {pages} {1 } (\bibinfo {year} {2018})},\ \bibinfo
  {note} {cascades and transitions in turbulent flows}\BibitemShut {NoStop}%
\bibitem [{\citenamefont {Benavides}\ and\ \citenamefont
  {Bustamante}(2025)}]{benavides:2025}%
  \BibitemOpen
  \bibfield  {author} {\bibinfo {author} {\bibfnamefont {S.~J.}\ \bibnamefont
  {Benavides}}\ and\ \bibinfo {author} {\bibfnamefont {M.~D.}\ \bibnamefont
  {Bustamante}},\ }\href {https://arxiv.org/abs/2507.03397} {\enquote {\bibinfo
  {title} {Phase dynamics and their role determining energy flux in
  hydrodynamic shell models},}\ } (\bibinfo {year} {2025}),\ \Eprint
  {https://arxiv.org/abs/2507.03397} {arXiv:2507.03397 [physics.flu-dyn]}
  \BibitemShut {NoStop}%
\bibitem [{\citenamefont {Biferale}\ \emph {et~al.}(1995)\citenamefont
  {Biferale}, \citenamefont {Lambert}, \citenamefont {Lima},\ and\
  \citenamefont {Paladin}}]{biferale:95}%
  \BibitemOpen
  \bibfield  {author} {\bibinfo {author} {\bibfnamefont {L.}~\bibnamefont
  {Biferale}}, \bibinfo {author} {\bibfnamefont {A.}~\bibnamefont {Lambert}},
  \bibinfo {author} {\bibfnamefont {R.}~\bibnamefont {Lima}},\ and\ \bibinfo
  {author} {\bibfnamefont {G.}~\bibnamefont {Paladin}},\ }\href@noop {}
  {\bibfield  {journal} {\bibinfo  {journal} {Physica D: Nonlinear Phenomena}\
  }\textbf {\bibinfo {volume} {80}},\ \bibinfo {pages} {105 } (\bibinfo {year}
  {1995})}\BibitemShut {NoStop}%
\bibitem [{\citenamefont {Ditlevsen}\ and\ \citenamefont
  {Mogensen}(1996)}]{ditlevsen:1996}%
  \BibitemOpen
  \bibfield  {author} {\bibinfo {author} {\bibfnamefont {P.}~\bibnamefont
  {Ditlevsen}}\ and\ \bibinfo {author} {\bibfnamefont {I.}~\bibnamefont
  {Mogensen}},\ }\href {https://doi.org/10.1103/PhysRevE.53.4785} {\bibfield
  {journal} {\bibinfo  {journal} {Phys. Rev. E}\ }\textbf {\bibinfo {volume}
  {53}},\ \bibinfo {pages} {4785} (\bibinfo {year} {1996})}\BibitemShut
  {NoStop}%
\bibitem [{\citenamefont {Rathmann}\ and\ \citenamefont
  {Ditlevsen}(2016)}]{rathmann}%
  \BibitemOpen
  \bibfield  {author} {\bibinfo {author} {\bibfnamefont {N.~M.}\ \bibnamefont
  {Rathmann}}\ and\ \bibinfo {author} {\bibfnamefont {P.~D.}\ \bibnamefont
  {Ditlevsen}},\ }\href {https://doi.org/10.1103/PhysRevE.94.033115} {\bibfield
   {journal} {\bibinfo  {journal} {Phys. Rev. E}\ }\textbf {\bibinfo {volume}
  {94}} (\bibinfo {year} {2016}),\ 10.1103/PhysRevE.94.033115}\BibitemShut
  {NoStop}%
\bibitem [{\citenamefont {Pisarenko}\ \emph {et~al.}(1993)\citenamefont
  {Pisarenko}, \citenamefont {Biferale}, \citenamefont {Courvoisier},
  \citenamefont {Frisch},\ and\ \citenamefont {Vergassola}}]{pisarenko:93}%
  \BibitemOpen
  \bibfield  {author} {\bibinfo {author} {\bibfnamefont {D.}~\bibnamefont
  {Pisarenko}}, \bibinfo {author} {\bibfnamefont {L.}~\bibnamefont {Biferale}},
  \bibinfo {author} {\bibfnamefont {D.}~\bibnamefont {Courvoisier}}, \bibinfo
  {author} {\bibfnamefont {U.}~\bibnamefont {Frisch}},\ and\ \bibinfo {author}
  {\bibfnamefont {M.}~\bibnamefont {Vergassola}},\ }\href
  {https://doi.org/10.1063/1.858766} {\bibfield  {journal} {\bibinfo  {journal}
  {Physics of Fluids A}\ }\textbf {\bibinfo {volume} {5}},\ \bibinfo {pages}
  {2533} (\bibinfo {year} {1993})}\BibitemShut {NoStop}%
\bibitem [{\citenamefont {Biferale}\ \emph {et~al.}(1999)\citenamefont
  {Biferale}, \citenamefont {Boffetta}, \citenamefont {Celani},\ and\
  \citenamefont {Toschi}}]{biferale1999}%
  \BibitemOpen
  \bibfield  {author} {\bibinfo {author} {\bibfnamefont {L.}~\bibnamefont
  {Biferale}}, \bibinfo {author} {\bibfnamefont {G.}~\bibnamefont {Boffetta}},
  \bibinfo {author} {\bibfnamefont {A.}~\bibnamefont {Celani}},\ and\ \bibinfo
  {author} {\bibfnamefont {F.}~\bibnamefont {Toschi}},\ }\href
  {https://doi.org/https://doi.org/10.1016/S0167-2789(98)00277-2} {\bibfield
  {journal} {\bibinfo  {journal} {Physica D: Nonlinear Phenomena}\ }\textbf
  {\bibinfo {volume} {127}},\ \bibinfo {pages} {187} (\bibinfo {year}
  {1999})}\BibitemShut {NoStop}%
\bibitem [{\citenamefont {Benzi}, \citenamefont {Biferale},\ and\ \citenamefont
  {Parisi}(1993)}]{benzi:93}%
  \BibitemOpen
  \bibfield  {author} {\bibinfo {author} {\bibfnamefont {R.}~\bibnamefont
  {Benzi}}, \bibinfo {author} {\bibfnamefont {L.}~\bibnamefont {Biferale}},\
  and\ \bibinfo {author} {\bibfnamefont {G.}~\bibnamefont {Parisi}},\ }\href
  {https://doi.org/10.1016/0167-2789(93)90012-P} {\bibfield  {journal}
  {\bibinfo  {journal} {Physica D: Nonlinear Phenomena}\ }\textbf {\bibinfo
  {volume} {65}},\ \bibinfo {pages} {163 } (\bibinfo {year}
  {1993})}\BibitemShut {NoStop}%
\bibitem [{\citenamefont {Biferale}, \citenamefont {Mailybaev},\ and\
  \citenamefont {Parisi}(2017)}]{biferale:2017}%
  \BibitemOpen
  \bibfield  {author} {\bibinfo {author} {\bibfnamefont {L.}~\bibnamefont
  {Biferale}}, \bibinfo {author} {\bibfnamefont {A.~A.}\ \bibnamefont
  {Mailybaev}},\ and\ \bibinfo {author} {\bibfnamefont {G.}~\bibnamefont
  {Parisi}},\ }\href {https://doi.org/10.1103/PhysRevE.95.043108} {\bibfield
  {journal} {\bibinfo  {journal} {PHYSICAL REVIEW E}\ }\textbf {\bibinfo
  {volume} {95}} (\bibinfo {year} {2017}),\
  10.1103/PhysRevE.95.043108}\BibitemShut {NoStop}%
\bibitem [{\citenamefont {Vladimirova}, \citenamefont {Shavit},\ and\
  \citenamefont {Falkovich}(2021)}]{vladimirova:2021}%
  \BibitemOpen
  \bibfield  {author} {\bibinfo {author} {\bibfnamefont {N.}~\bibnamefont
  {Vladimirova}}, \bibinfo {author} {\bibfnamefont {M.}~\bibnamefont
  {Shavit}},\ and\ \bibinfo {author} {\bibfnamefont {G.}~\bibnamefont
  {Falkovich}},\ }\href {https://doi.org/10.1103/PhysRevX.11.021063} {\bibfield
   {journal} {\bibinfo  {journal} {Phys. Rev. X}\ }\textbf {\bibinfo {volume}
  {11}},\ \bibinfo {pages} {021063} (\bibinfo {year} {2021})}\BibitemShut
  {NoStop}%
\bibitem [{\citenamefont {Strogatz}(2001)}]{strogatz:01}%
  \BibitemOpen
  \bibfield  {author} {\bibinfo {author} {\bibfnamefont {S.~H.}\ \bibnamefont
  {Strogatz}},\ }\href@noop {} {\bibfield  {journal} {\bibinfo  {journal}
  {Nature}\ }\textbf {\bibinfo {volume} {410}},\ \bibinfo {pages} {268}
  (\bibinfo {year} {2001})}\BibitemShut {NoStop}%
\bibitem [{\citenamefont {Strogatz}, \citenamefont {Mirollo},\ and\
  \citenamefont {Matthews}(1992)}]{strogatz:92}%
  \BibitemOpen
  \bibfield  {author} {\bibinfo {author} {\bibfnamefont {S.~H.}\ \bibnamefont
  {Strogatz}}, \bibinfo {author} {\bibfnamefont {R.~E.}\ \bibnamefont
  {Mirollo}},\ and\ \bibinfo {author} {\bibfnamefont {P.~C.}\ \bibnamefont
  {Matthews}},\ }\href@noop {} {\bibfield  {journal} {\bibinfo  {journal}
  {Phys. Rev. Lett.}\ }\textbf {\bibinfo {volume} {68}},\ \bibinfo {pages}
  {2730} (\bibinfo {year} {1992})}\BibitemShut {NoStop}%
\bibitem [{\citenamefont {Kuramoto}(1984)}]{kuramoto:book:1984}%
  \BibitemOpen
  \bibfield  {author} {\bibinfo {author} {\bibfnamefont {Y.}~\bibnamefont
  {Kuramoto}},\ }\href@noop {} {\emph {\bibinfo {title} {{Chemical
  Oscillations, Waves, and Turbulence}}}}\ (\bibinfo  {publisher}
  {Springer--Verlag},\ \bibinfo {address} {New York},\ \bibinfo {year}
  {1984})\BibitemShut {NoStop}%
\bibitem [{\citenamefont {Tom}\ and\ \citenamefont {Ray}(2017)}]{tom:2017}%
  \BibitemOpen
  \bibfield  {author} {\bibinfo {author} {\bibfnamefont {R.}~\bibnamefont
  {Tom}}\ and\ \bibinfo {author} {\bibfnamefont {S.~S.}\ \bibnamefont {Ray}},\
  }\href {https://doi.org/10.1209/0295-5075/120/34002} {\bibfield  {journal}
  {\bibinfo  {journal} {EPL}\ }\textbf {\bibinfo {volume} {120}} (\bibinfo
  {year} {2017}),\ 10.1209/0295-5075/120/34002}\BibitemShut {NoStop}%
\bibitem [{\citenamefont {Arguedas-Leiva}\ \emph {et~al.}(2022)\citenamefont
  {Arguedas-Leiva}, \citenamefont {Carroll}, \citenamefont {Biferale},
  \citenamefont {Wilczek},\ and\ \citenamefont {Bustamante}}]{leiva:2022}%
  \BibitemOpen
  \bibfield  {author} {\bibinfo {author} {\bibfnamefont {J.-A.}\ \bibnamefont
  {Arguedas-Leiva}}, \bibinfo {author} {\bibfnamefont {E.}~\bibnamefont
  {Carroll}}, \bibinfo {author} {\bibfnamefont {L.}~\bibnamefont {Biferale}},
  \bibinfo {author} {\bibfnamefont {M.}~\bibnamefont {Wilczek}},\ and\ \bibinfo
  {author} {\bibfnamefont {M.~D.}\ \bibnamefont {Bustamante}},\ }\href
  {https://doi.org/10.1103/PhysRevResearch.4.L032035} {\bibfield  {journal}
  {\bibinfo  {journal} {Phys. Rev. Res.}\ }\textbf {\bibinfo {volume} {4}},\
  \bibinfo {pages} {L032035} (\bibinfo {year} {2022})}\BibitemShut {NoStop}%
\bibitem [{\citenamefont {Benzi}\ \emph {et~al.}(1993)\citenamefont {Benzi},
  \citenamefont {Ciliberto}, \citenamefont {Tripiccione}, \citenamefont
  {Baudet}, \citenamefont {Massaioli},\ and\ \citenamefont {Succi}}]{benzi93}%
  \BibitemOpen
  \bibfield  {author} {\bibinfo {author} {\bibfnamefont {R.}~\bibnamefont
  {Benzi}}, \bibinfo {author} {\bibfnamefont {S.}~\bibnamefont {Ciliberto}},
  \bibinfo {author} {\bibfnamefont {R.}~\bibnamefont {Tripiccione}}, \bibinfo
  {author} {\bibfnamefont {C.}~\bibnamefont {Baudet}}, \bibinfo {author}
  {\bibfnamefont {F.}~\bibnamefont {Massaioli}},\ and\ \bibinfo {author}
  {\bibfnamefont {S.}~\bibnamefont {Succi}},\ }\href
  {https://doi.org/10.1103/PhysRevE.48.R29} {\bibfield  {journal} {\bibinfo
  {journal} {Phys. Rev. E}\ }\textbf {\bibinfo {volume} {48}},\ \bibinfo
  {pages} {R29} (\bibinfo {year} {1993})}\BibitemShut {NoStop}%
\bibitem [{\citenamefont {Banerjee}\ \emph {et~al.}(2013)\citenamefont
  {Banerjee}, \citenamefont {Ray}, \citenamefont {Sahoo},\ and\ \citenamefont
  {Pandit}}]{banerjee:2013}%
  \BibitemOpen
  \bibfield  {author} {\bibinfo {author} {\bibfnamefont {D.}~\bibnamefont
  {Banerjee}}, \bibinfo {author} {\bibfnamefont {S.~S.}\ \bibnamefont {Ray}},
  \bibinfo {author} {\bibfnamefont {G.}~\bibnamefont {Sahoo}},\ and\ \bibinfo
  {author} {\bibfnamefont {R.}~\bibnamefont {Pandit}},\ }\href
  {https://doi.org/10.1103/PhysRevLett.111.174501} {\bibfield  {journal}
  {\bibinfo  {journal} {Phys. Rev. Lett.}\ }\textbf {\bibinfo {volume} {111}},\
  \bibinfo {pages} {174501} (\bibinfo {year} {2013})}\BibitemShut {NoStop}%
\bibitem [{\citenamefont {Chakraborty}, \citenamefont {Frisch},\ and\
  \citenamefont {Ray}(2010)}]{chakraborty:2010}%
  \BibitemOpen
  \bibfield  {author} {\bibinfo {author} {\bibfnamefont {S.}~\bibnamefont
  {Chakraborty}}, \bibinfo {author} {\bibfnamefont {U.}~\bibnamefont
  {Frisch}},\ and\ \bibinfo {author} {\bibfnamefont {S.~S.}\ \bibnamefont
  {Ray}},\ }\href {https://doi.org/10.1017/S0022112010000595} {\bibfield
  {journal} {\bibinfo  {journal} {J. Fluid Mech.}\ }\textbf {\bibinfo {volume}
  {649}},\ \bibinfo {pages} {275} (\bibinfo {year} {2010})}\BibitemShut
  {NoStop}%
\bibitem [{\citenamefont {Benzi}, \citenamefont {Biferale},\ and\ \citenamefont
  {Trovatore}(1996)}]{benzi:1996}%
  \BibitemOpen
  \bibfield  {author} {\bibinfo {author} {\bibfnamefont {R.}~\bibnamefont
  {Benzi}}, \bibinfo {author} {\bibfnamefont {L.}~\bibnamefont {Biferale}},\
  and\ \bibinfo {author} {\bibfnamefont {E.}~\bibnamefont {Trovatore}},\ }\href
  {https://doi.org/10.1103/PhysRevLett.77.3114} {\bibfield  {journal} {\bibinfo
   {journal} {Phys. Rev. Lett.}\ }\textbf {\bibinfo {volume} {77}},\ \bibinfo
  {pages} {3114} (\bibinfo {year} {1996})}\BibitemShut {NoStop}%
\bibitem [{\citenamefont {Andersen}\ \emph {et~al.}(2000)\citenamefont
  {Andersen}, \citenamefont {Bohr}, \citenamefont {Jensen}, \citenamefont
  {Nielsen},\ and\ \citenamefont {Olesen}}]{Andersen:00}%
  \BibitemOpen
  \bibfield  {author} {\bibinfo {author} {\bibfnamefont {K.}~\bibnamefont
  {Andersen}}, \bibinfo {author} {\bibfnamefont {T.}~\bibnamefont {Bohr}},
  \bibinfo {author} {\bibfnamefont {M.}~\bibnamefont {Jensen}}, \bibinfo
  {author} {\bibfnamefont {J.}~\bibnamefont {Nielsen}},\ and\ \bibinfo {author}
  {\bibfnamefont {P.}~\bibnamefont {Olesen}},\ }\href@noop {} {\bibfield
  {journal} {\bibinfo  {journal} {Physica D}\ }\textbf {\bibinfo {volume}
  {138}},\ \bibinfo {pages} {44} (\bibinfo {year} {2000})}\BibitemShut
  {NoStop}%
\bibitem [{\citenamefont {Bec}\ and\ \citenamefont {Khanin}(2007)}]{bec:2007}%
  \BibitemOpen
  \bibfield  {author} {\bibinfo {author} {\bibfnamefont {J.}~\bibnamefont
  {Bec}}\ and\ \bibinfo {author} {\bibfnamefont {K.}~\bibnamefont {Khanin}},\
  }\href {https://doi.org/https://doi.org/10.1016/j.physrep.2007.04.002}
  {\bibfield  {journal} {\bibinfo  {journal} {Physics Reports}\ }\textbf
  {\bibinfo {volume} {447}},\ \bibinfo {pages} {1} (\bibinfo {year}
  {2007})}\BibitemShut {NoStop}%
\bibitem [{\citenamefont {Benzi}\ \emph {et~al.}(2022)\citenamefont {Benzi},
  \citenamefont {Castaldi}, \citenamefont {Toschi},\ and\ \citenamefont
  {Trampert}}]{benzi:2022}%
  \BibitemOpen
  \bibfield  {author} {\bibinfo {author} {\bibfnamefont {R.}~\bibnamefont
  {Benzi}}, \bibinfo {author} {\bibfnamefont {I.}~\bibnamefont {Castaldi}},
  \bibinfo {author} {\bibfnamefont {F.}~\bibnamefont {Toschi}},\ and\ \bibinfo
  {author} {\bibfnamefont {J.}~\bibnamefont {Trampert}},\ }\href
  {https://doi.org/10.1098/rsta.2021.0074} {\bibfield  {journal} {\bibinfo
  {journal} {Phil. Trans. R. Soc. A}\ }\textbf {\bibinfo {volume} {380}},\
  \bibinfo {pages} {20210074} (\bibinfo {year} {2022})},\ \Eprint
  {https://arxiv.org/abs/https://royalsocietypublishing.org/doi/pdf/10.1098/rsta.2021.0074}
  {https://royalsocietypublishing.org/doi/pdf/10.1098/rsta.2021.0074}
  \BibitemShut {NoStop}%
\bibitem [{\citenamefont {Waleffe}(1992)}]{waleffe:92}%
  \BibitemOpen
  \bibfield  {author} {\bibinfo {author} {\bibfnamefont {F.}~\bibnamefont
  {Waleffe}},\ }\href {https://doi.org/10.1063/1.858309} {\bibfield  {journal}
  {\bibinfo  {journal} {Physics of Fluids A: Fluid Dynamics}\ }\textbf
  {\bibinfo {volume} {4}},\ \bibinfo {pages} {350} (\bibinfo {year}
  {1992})}\BibitemShut {NoStop}%
\bibitem [{\citenamefont {Benzi}\ \emph {et~al.}(1996)\citenamefont {Benzi},
  \citenamefont {Biferale}, \citenamefont {Kerr},\ and\ \citenamefont
  {Trovatore}}]{benzi:96}%
  \BibitemOpen
  \bibfield  {author} {\bibinfo {author} {\bibfnamefont {R.}~\bibnamefont
  {Benzi}}, \bibinfo {author} {\bibfnamefont {L.}~\bibnamefont {Biferale}},
  \bibinfo {author} {\bibfnamefont {R.~M.}\ \bibnamefont {Kerr}},\ and\
  \bibinfo {author} {\bibfnamefont {E.}~\bibnamefont {Trovatore}},\ }\href@noop
  {} {\bibfield  {journal} {\bibinfo  {journal} {Phys. Rev. E}\ }\textbf
  {\bibinfo {volume} {53}},\ \bibinfo {pages} {3541} (\bibinfo {year}
  {1996})}\BibitemShut {NoStop}%
\bibitem [{\citenamefont {De~Pietro}, \citenamefont {Biferale},\ and\
  \citenamefont {Mailybaev}(2015)}]{depietro:15}%
  \BibitemOpen
  \bibfield  {author} {\bibinfo {author} {\bibfnamefont {M.}~\bibnamefont
  {De~Pietro}}, \bibinfo {author} {\bibfnamefont {L.}~\bibnamefont
  {Biferale}},\ and\ \bibinfo {author} {\bibfnamefont {A.~A.}\ \bibnamefont
  {Mailybaev}},\ }\href@noop {} {\bibfield  {journal} {\bibinfo  {journal}
  {Phys. Rev. E}\ }\textbf {\bibinfo {volume} {92}},\ \bibinfo {pages} {043021}
  (\bibinfo {year} {2015})}\BibitemShut {NoStop}%
\bibitem [{\citenamefont {De~Pietro}, \citenamefont {Mailybaev},\ and\
  \citenamefont {Biferale}(2017)}]{depietro:2017}%
  \BibitemOpen
  \bibfield  {author} {\bibinfo {author} {\bibfnamefont {M.}~\bibnamefont
  {De~Pietro}}, \bibinfo {author} {\bibfnamefont {A.~A.}\ \bibnamefont
  {Mailybaev}},\ and\ \bibinfo {author} {\bibfnamefont {L.}~\bibnamefont
  {Biferale}},\ }\href {https://doi.org/10.1103/PhysRevFluids.2.034606}
  {\bibfield  {journal} {\bibinfo  {journal} {Phys. Rev. Fluids}\ }\textbf
  {\bibinfo {volume} {2}},\ \bibinfo {pages} {034606} (\bibinfo {year}
  {2017})}\BibitemShut {NoStop}%
\bibitem [{\citenamefont {Aurell}\ \emph {et~al.}(1994)\citenamefont {Aurell},
  \citenamefont {Boffetta}, \citenamefont {Crisanti}, \citenamefont {Frick},
  \citenamefont {Paladin},\ and\ \citenamefont {Vulpiani}}]{aurell:94b}%
  \BibitemOpen
  \bibfield  {author} {\bibinfo {author} {\bibfnamefont {E.}~\bibnamefont
  {Aurell}}, \bibinfo {author} {\bibfnamefont {G.}~\bibnamefont {Boffetta}},
  \bibinfo {author} {\bibfnamefont {A.}~\bibnamefont {Crisanti}}, \bibinfo
  {author} {\bibfnamefont {P.}~\bibnamefont {Frick}}, \bibinfo {author}
  {\bibfnamefont {G.}~\bibnamefont {Paladin}},\ and\ \bibinfo {author}
  {\bibfnamefont {A.}~\bibnamefont {Vulpiani}},\ }\href
  {https://doi.org/10.1103/PhysRevE.50.4705} {\bibfield  {journal} {\bibinfo
  {journal} {Phys. Rev. E}\ }\textbf {\bibinfo {volume} {50}},\ \bibinfo
  {pages} {4705} (\bibinfo {year} {1994})}\BibitemShut {NoStop}%
\bibitem [{\citenamefont {Aurell}, \citenamefont {Frick},\ and\ \citenamefont
  {Shaidurov}(1994)}]{aurell:94}%
  \BibitemOpen
  \bibfield  {author} {\bibinfo {author} {\bibfnamefont {E.}~\bibnamefont
  {Aurell}}, \bibinfo {author} {\bibfnamefont {P.}~\bibnamefont {Frick}},\ and\
  \bibinfo {author} {\bibfnamefont {V.}~\bibnamefont {Shaidurov}},\ }\href@noop
  {} {\bibfield  {journal} {\bibinfo  {journal} {Physica D: Nonlinear
  Phenomena}\ }\textbf {\bibinfo {volume} {72}},\ \bibinfo {pages} {95 }
  (\bibinfo {year} {1994})}\BibitemShut {NoStop}%
\bibitem [{\citenamefont {Aurell}, \citenamefont {Dormy},\ and\ \citenamefont
  {Frick}(1997)}]{aurell:97}%
  \BibitemOpen
  \bibfield  {author} {\bibinfo {author} {\bibfnamefont {E.}~\bibnamefont
  {Aurell}}, \bibinfo {author} {\bibfnamefont {E.}~\bibnamefont {Dormy}},\ and\
  \bibinfo {author} {\bibfnamefont {P.}~\bibnamefont {Frick}},\ }\href@noop {}
  {\bibfield  {journal} {\bibinfo  {journal} {Phys. Rev. E}\ }\textbf {\bibinfo
  {volume} {56}},\ \bibinfo {pages} {1692} (\bibinfo {year}
  {1997})}\BibitemShut {NoStop}%
\bibitem [{\citenamefont {Ditlevsen}(2012)}]{ditlevsen:12}%
  \BibitemOpen
  \bibfield  {author} {\bibinfo {author} {\bibfnamefont {P.~D.}\ \bibnamefont
  {Ditlevsen}},\ }\href {https://doi.org/10.1063/1.4761834} {\bibfield
  {journal} {\bibinfo  {journal} {Physics of Fluids}\ }\textbf {\bibinfo
  {volume} {24}},\ \bibinfo {pages} {105109} (\bibinfo {year}
  {2012})}\BibitemShut {NoStop}%
\bibitem [{\citenamefont {Rathmann}\ and\ \citenamefont
  {Ditlevsen}(2017)}]{rathmann:2017}%
  \BibitemOpen
  \bibfield  {author} {\bibinfo {author} {\bibfnamefont {N.~M.}\ \bibnamefont
  {Rathmann}}\ and\ \bibinfo {author} {\bibfnamefont {P.~D.}\ \bibnamefont
  {Ditlevsen}},\ }\href {https://doi.org/10.1103/PhysRevFluids.2.054607}
  {\bibfield  {journal} {\bibinfo  {journal} {Phys. Rev. Fluids}\ }\textbf
  {\bibinfo {volume} {2}},\ \bibinfo {pages} {054607} (\bibinfo {year}
  {2017})}\BibitemShut {NoStop}%
\bibitem [{\citenamefont {Gilbert}\ \emph {et~al.}(2002)\citenamefont
  {Gilbert}, \citenamefont {L'vov}, \citenamefont {Pomyalov},\ and\
  \citenamefont {Procaccia}}]{gilbert:2002}%
  \BibitemOpen
  \bibfield  {author} {\bibinfo {author} {\bibfnamefont {T.}~\bibnamefont
  {Gilbert}}, \bibinfo {author} {\bibfnamefont {V.}~\bibnamefont {L'vov}},
  \bibinfo {author} {\bibfnamefont {A.}~\bibnamefont {Pomyalov}},\ and\
  \bibinfo {author} {\bibfnamefont {I.}~\bibnamefont {Procaccia}},\ }\href
  {https://doi.org/10.1103/PhysRevLett.89.074501} {\bibfield  {journal}
  {\bibinfo  {journal} {Phys. Rev. Lett.}\ }\textbf {\bibinfo {volume} {89}}
  (\bibinfo {year} {2002}),\ 10.1103/PhysRevLett.89.074501}\BibitemShut
  {NoStop}%
\bibitem [{\citenamefont {Mailybaev}(2022{\natexlab{a}})}]{mailybaev:2022}%
  \BibitemOpen
  \bibfield  {author} {\bibinfo {author} {\bibfnamefont {A.~A.}\ \bibnamefont
  {Mailybaev}},\ }\href {https://doi.org/10.1103/PhysRevFluids.7.034604}
  {\bibfield  {journal} {\bibinfo  {journal} {Phys. Rev. Fluids}\ }\textbf
  {\bibinfo {volume} {7}},\ \bibinfo {pages} {034604} (\bibinfo {year}
  {2022}{\natexlab{a}})}\BibitemShut {NoStop}%
\bibitem [{\citenamefont {Mailybaev}(2022{\natexlab{b}})}]{mailybaev_rev:2022}%
  \BibitemOpen
  \bibfield  {author} {\bibinfo {author} {\bibfnamefont {A.~A.}\ \bibnamefont
  {Mailybaev}},\ }\href {https://doi.org/10.1088/1361-6544/ac7504} {\bibfield
  {journal} {\bibinfo  {journal} {Nonlinearity}\ }\textbf {\bibinfo {volume}
  {35}},\ \bibinfo {pages} {3630} (\bibinfo {year}
  {2022}{\natexlab{b}})}\BibitemShut {NoStop}%
\bibitem [{\citenamefont {Manfredini}\ and\ \citenamefont
  {G{\"u}rcan}(2025)}]{manfredinidataset:2025}%
  \BibitemOpen
  \bibfield  {author} {\bibinfo {author} {\bibfnamefont {L.}~\bibnamefont
  {Manfredini}}\ and\ \bibinfo {author} {\bibfnamefont {{\"O}.}~\bibnamefont
  {G{\"u}rcan}},\ }\href {https://doi.org/10.5281/zenodo.17514728} {\enquote
  {\bibinfo {title} {Repository for nonlinear phase synchronization and the
  role of spacing in shell models},}\ } (\bibinfo {year} {2025}),\ \bibinfo
  {note} {zenodo dataset}\BibitemShut {NoStop}%
\bibitem [{\citenamefont {Rackauckas}(2024)}]{rackauckas24}%
  \BibitemOpen
  \bibfield  {author} {\bibinfo {author} {\bibfnamefont {C.}~\bibnamefont
  {Rackauckas}},\ }\href {https://doi.org/10.5281/zenodo.13629099} {\enquote
  {\bibinfo {title} {Sciml/differentialequations.jl: v7.14.0},}\ }\bibinfo
  {howpublished} {https://doi.org/10.5281/zenodo.13629099} (\bibinfo {year}
  {2024})\BibitemShut {NoStop}%
\bibitem [{\citenamefont {Kennedy}\ and\ \citenamefont
  {Carpenter}(2003)}]{kennedy2003AdditiveRS}%
  \BibitemOpen
  \bibfield  {author} {\bibinfo {author} {\bibfnamefont {C.~A.}\ \bibnamefont
  {Kennedy}}\ and\ \bibinfo {author} {\bibfnamefont {M.~H.}\ \bibnamefont
  {Carpenter}},\ }\href {https://api.semanticscholar.org/CorpusID:4685712}
  {\bibfield  {journal} {\bibinfo  {journal} {Applied Numerical Mathematics}\
  }\textbf {\bibinfo {volume} {44}},\ \bibinfo {pages} {139} (\bibinfo {year}
  {2003})}\BibitemShut {NoStop}%
\bibitem [{\citenamefont {Bezanson}\ \emph {et~al.}(2017)\citenamefont
  {Bezanson}, \citenamefont {Edelman}, \citenamefont {Karpinski},\ and\
  \citenamefont {Shah}}]{julia2017}%
  \BibitemOpen
  \bibfield  {author} {\bibinfo {author} {\bibfnamefont {J.}~\bibnamefont
  {Bezanson}}, \bibinfo {author} {\bibfnamefont {A.}~\bibnamefont {Edelman}},
  \bibinfo {author} {\bibfnamefont {S.}~\bibnamefont {Karpinski}},\ and\
  \bibinfo {author} {\bibfnamefont {V.~B.}\ \bibnamefont {Shah}},\ }\href
  {https://doi.org/10.1137/141000671} {\bibfield  {journal} {\bibinfo
  {journal} {SIAM Review}\ }\textbf {\bibinfo {volume} {59}},\ \bibinfo {pages}
  {65} (\bibinfo {year} {2017})},\ \Eprint
  {https://arxiv.org/abs/https://doi.org/10.1137/141000671}
  {https://doi.org/10.1137/141000671} \BibitemShut {NoStop}%
\end{thebibliography}
%aipnum4-2.bst 2019-01-14 (MD) hand-edited version of apsrev4-1.bst

\appendix

\section{\protect\label{sec:Numerical-details}Numerical details}

The dynamics of the different shell models presented have been integrated
using a $4$th order Implicit-Explicit (IMEX) solver \citep{rackauckas24,kennedy2003AdditiveRS}
implemented in Julia \citep{julia2017}. This approach allows for
efficient separation of the stiff and non-stiff components in the
coupled ordinary differential equations system that govern the shell
model evolution.

For all the shell models considered, given the inter-shell spacing
$g$, we fix a number of shells $N$ in order to cover approximately
seven decades. The dissipation term is set as $d_{n}=\nu k_{n}^{2}$
where the viscosity $\nu$ is chosen small enough to dissipate energy
only in the last few shells. For the inverse cascade case, an additional
hypo-viscosity term ($\mu k_{n}^{-2}$) has been included to prevent
energy accumulation at large scales.

The statistical analyses are performed in a steady state regime, over
a time window corresponding to approximately 2,000 eddy turnover times
($\tau_{E}^{-1}\sim k_{n}u_{n}$) of the largest scale. The sampling
time is set to $\delta t=5\times10^{-5}$, ensuring a sufficiently
high sampling rate to resolve the temporal dynamics within the inertial
range. The initial condition typically consists of a small amplitude
$(10^{-8})$ white noise. The Tables \ref{tab:goy-params-1}, \ref{tab:params-helical-1}
and \ref{tab:params-inv-1} summarize all the parameters used for
the various runs that corresponding to the numerical experiments analyzed
in the paper.

A special note concerns the choice of the forcing term. The common
approach in the literature is to apply either a constant or random
forcing. In the specific case discussed in Sec. \ref{sec:Intermittency-1}
to investigate the role of phase self-organization, particularly in
relation to phase-syncronized events approaching the continuum limit,
we implemented a forcing term that does not inject any phase into
the system. This is achieved by ensuring that the term $Im\left(\frac{f_{n}}{\rho_{n}}e^{-i\theta_{n}}\right)$
on the right-hand side of Eq. (\ref{eq:phgoy}) is identically zero,
thus we fixed $f_{n}=Ae^{i\theta_{n}}$, with $A=0.5$ , in all the
runs corresponding to Table \ref{tab:goy-params-1}. Another possible
choice, $f_{n}=\epsilon/u_{n}^{*}$, which does not inject any phase
and fix a constant energy injection $\epsilon$, as discussed in Ref.
\citealp{benzi:2022} in order to study the statistics of inter-event
times. We note that for the result of present analysis there are no
significant differences among these two choices.

\begin{table*}[t]
\begin{tabular}{cccccccccccc}
\multicolumn{12}{c}{GOY model Runs}\tabularnewline
\hline 
\hline 
$g$ & 1.1 & 1.15 & 1.2 & 1.3 & 1.4 & 1.5 & 1.6 & 1.7 & 1.8 & 1.9 & 2.0\tabularnewline
\hline 
$N$ & 164 & 113 & 88 & 62 & 50 & 42 & 38 & 36 & 33 & 31 & 29\tabularnewline
\hline 
$\nu$ & $5\times10^{-9}$ & $4\times10^{-9}$ & $8\times10^{-9}$ & $14\times10^{-9}$ & $9\times10^{-9}$ & $8\times10^{-9}$ & $5\times10^{-9}$ & $9\times10^{-10}$ & $7\times10^{-10}$ & $5.5\times10^{-10}$ & $5.5\times10^{-10}$\tabularnewline
\end{tabular}

\caption{\protect\label{tab:goy-params-1}Numerical parameters corresponding
to the different runs of the GOY model presented is Sec. \ref{sec:Intermittency-1}.
In all the different runs the shell wave-numbers correspond to $k_{n}=k_{0}g^{n}$
with $k_{0}=1/g^{3}$, and the forcing act the third shell i.e. $f_{3}=Ae^{i\theta_{3}}$
with $A=0.5$.}
\end{table*}

\begin{table*}[t]
\begin{tabular}{cccccccc}
\multicolumn{8}{c}{Helical shell models Runs}\tabularnewline
\hline 
\hline 
class & SM1 & SM1 & SM2 & SM2 & SM3 & SM3 & SM3\tabularnewline
\hline 
$g$ & 1.1 & 2 & 1.1 & 2 & 1.1 & 1.2 & 2\tabularnewline
\hline 
$N$ & 164 & 30 & 140 & 29 & 140 & 80 & 30\tabularnewline
\hline 
$\nu$ & $5\times10^{-10}$ & $1.5\times10^{-10}$ & $8\times10^{-10}$ & $3.5\times10^{-10}$ & $8\times10^{-10}$ & $1\times10^{-9}$ & $1.5\times10^{-10}$\tabularnewline
\end{tabular}

\caption{\protect\label{tab:params-helical-1}Numerical parameters corresponding
to the different runs of the local Helical shell models.$f_{3}=(1+1j)\times7\times10^{-1}$}
\end{table*}

\begin{table*}[t]
\begin{tabular}{ccc}
\multicolumn{3}{c}{Inverse cascade shell models Runs}\tabularnewline
\hline 
\hline 
model & Helical-nonlocal & Sabra-local\tabularnewline
\hline 
$g$ & $\varphi$ & 2\tabularnewline
\hline 
$N$ & 36 & 28\tabularnewline
\hline 
$\nu$ & $4\times10^{-12}$ & $5\times10^{-12}$\tabularnewline
\hline 
$\mu$ & 1 & $1\times10^{-5}$\tabularnewline
\hline 
$f_{n}$ & $f_{32,33}^{+}=0.8$ , $f_{32,33}^{-}=0.4$ & $f_{23}=(5+5j)\times10^{-4}$\tabularnewline
\end{tabular}

\caption{\protect\label{tab:params-inv-1}Numerical parameters corresponding
to the different runs of the non local Helical shell model and the
local sabra model discussed in Sec.\ref{subsec:inverse-ray-model}}
\end{table*}

\end{document}